\documentclass[twocolumn,floatfix,aps,prb,eqsecnum,superscriptaddress]{revtex4}
\usepackage{tabularx,graphicx}
\usepackage{amsmath}
\usepackage{bm}
\usepackage{graphicx}
\usepackage[english]{babel}
\usepackage[version=3]{mhchem}	
\usepackage{siunitx}
\usepackage{setspace}
\usepackage{miller}
\usepackage{mathtools}
\usepackage{color}
\usepackage[FIGTOPCAP]{subfigure}
\usepackage{hyperref}

\def\k{{\bm k}}

\begin{document}

\title{Semiclassical theory of anisotropic transport at \ce{LaAlO3}/\ce{SrTiO3} interfaces under in-plane magnetic field}
\author{N.~Bovenzi}
\affiliation{Instituut-Lorentz, Universiteit Leiden, P.O. Box 9506, 2300 RA Leiden, The Netherlands}
\author{M.~Diez}
\affiliation{Instituut-Lorentz, Universiteit Leiden, P.O. Box 9506, 2300 RA Leiden, The Netherlands}
\date{September 2016}

\begin{abstract}

The unconventional magnetotransport at the interface between transition-metal oxides \ce{LaAlO3}~(\ce{LAO}) and \ce{SrTiO3} (\ce{STO}) is frequently related to mobile electrons interacting with localized magnetic moments. However nature and properties of magnetism at this interface are not well understood so far. In this paper, we focus on transport effects driven by spin-orbit coupling and intentionally neglect possible strong correlations. The electrical resistivity tensor is calculated as a function of the magnitude and orientation of an external magnetic field parallel to the interface. The semiclassical Boltzmann equation is solved numerically for the two-dimensional system of spin-orbit coupled electrons accelerated by an electric field and scattered by spatially-correlated impurities. At temperatures of a few Kelvin and densities such that the chemical potential crosses the second pair of spin-orbit split bands, we find a strongly anisotropic modulation of the (negative) magnetoresistance above $10\,{\rm T}$, characterized by multiple maxima and minima away from the crystalline axes. Along with the drop of the magnetoresistance, an abrupt enhancement of the transverse resistivity occurs. The angular modulation of the latter considerably deviates from a (low-field) sinusoidal dependence to a  (high-field) step-like behaviour. These peculiar features are the consequences of the anisotropy of both (\emph{intra-band} and \emph{inter-band}) scattering-amplitudes in the Brillouin zone when the relevant energy scales in the system -- chemical potential, spin-orbit interaction and Zeeman energy -- are all comparable to each other. The theory provides good qualitative agreement with experimental data in the literature.
\end{abstract}

\maketitle

Transition-metal oxide interfaces play a leading role in the development of \emph{quantum-matter heterostructures}, where novel electronic states are achievable due to the combination of the capabilities and rich variety of heterostructure engineering, the collective interactions of complex oxides, and the emergent properties of quantum materials\cite{boschker2016, mannhart2008}. 
A prototype system in this field is the heterostructure formed by the perovskite oxides \ce{LaAlO3} and \ce{SrTiO3}.
Since the experimental demonstration \cite{ohtomo2004} of electrical conduction at the interface between these two materials, large attention has been drawn to this system in particular due to its gate-tunable superconductivity \cite{reyren2007, caviglia2008, benshalom2010} at $T  \lesssim 300\,\si{\milli\kelvin}$.
At slightly higher temperatures - in the range $1-20\,\si{\kelvin}$ - magnetotransport has been an important tool for the investigation of electronic and magnetic properties of the interface that are believed to be strongly determined by mixing of charge, spin, orbital and lattice degrees of freedoms.
A number of signatures in the normal-state transport\cite{brinkman2007, benshalom2009, joshua2013, annadi2013}, such as giant negative magnetoresistance, crystalline anisotropy, anomalous Hall effect and their striking change of behavior when the system is tuned across a Lifshitz transition \cite{joshua2012} have been considered as an evidence of magnetism at the interface. In particular, Ruhmann et al.\cite{ruhman2014} suggested that the action of the field on the interaction between conduction electrons and localized magnetic moments induces a phase transition from a Kondo-screened (high and isotropic resistance) phase to a (low and anisotropic resistance) polarized phase, where the unscreened moments act as magnetic scatterers.

However, experimental investigations of the magnetic landscape at the interface \cite{ariando2011,
li2011, bert2011, lee2013, fitz2011, salman2012} reported qualitatively different results.
A strong ferromagnetic phase with large total magnetization was recently observed by magnetic force microscopy \cite{bi2014} at room-temperature in the depleted (insulating) regime of top-gated interfaces. On the other hand, the total magnetization was found to disappear when the interface was doped enough to be conducting.
More questions about the origin and the nature of magnetic structures at low temperatures remain to be answered.

Electrons at the interface are subjected to a spin-orbit interaction resulting from the interplay of the intrinsic spin-to-angular momentum coupling of $d$ atomic orbitals and the additional inter-orbital coupling activated by inversion-symmetry breaking at the interface. Experiments show that the effective spin-orbit coupling ({\rm SOC}) is highly tunable by means of electrostatic gating: an increase of one order of magnitude in the spin splitting at the Fermi level is estimated in the overdoped regime. \cite{caviglia2010, benshalom2010, fete2012} Spin-orbit interaction at the interface between complex-oxides has also been suggested as a possible source of electronic phase separation\cite{caprara2012, bucheli2013} and recently is the target of intensive effort in order to achieve room temperature spin-charge conversion and generating spin currents.\cite{song2016} 
Although commonly acknowledged as an important property of the system, only very recently it has been 
shown to have dramatic effects on the magnetotransport at the interface.\cite{diez2015}
Semiclassical calculations of the behaviour of the electrical resistivity with an external magnetic field parallel to the interface reproduced a large drop (up to $50-60 \%$) occurring on a field-scale of a few ${\rm T}$ due to the field-induced suppression of the \emph{inter-band} scattering. Importantly, the model also suggests a simple explanation to the striking similarity between gate-voltage and temperature dependence of the magnetoresistance -- revealed by the experimental data shown in the same paper \cite{diez2015} -- only in terms of properties of the band-structure and scattering by correlated impurities. 

In this work we apply the semiclassical model to study how the resistivity tensor evolves as a function of magnitude and direction of the in-plane magnetic field. The electrical response is found to be strongly anisotropic when the chemical potential falls into a range of the spectrum where electrons populate multiple subbands, with very different dispersions and spin-orbital structures in momentum-space. 
Furthermore, the anisotropy is characterized by a peculiar crossover from moderate to high fields, an explanation of which requires to consider the angular dependence of both \emph{inter-band} and \emph{intra-band} scattering.

The structure of the manuscipt is the following. In Sec.~\ref{experimental_review} we summarize the results of two experiments on \ce{LaAlO3}/\ce{SrTiO3} interfaces that systematically investigate the anisotropy of magnetotransport. This is not meant to be a complete review of the wide literature concerning this topic. In Sec.~\ref{theory_model} we introduce the theoretical model used for the numerical calculations. Starting from a low-energy non-interacting description of the two-dimensional electron system at the interface\cite{ruhman2014} (details in Appendix~\ref{model}), we focus on the structure of the spin-orbit coupling at the relevant energy scales. 
Numerical results are obtained by solving the Boltzmann transport equation in the weak-disorder limit, for two different models of spatially-correlated impurities.\cite{pikulin2011, fu2016}
In Sec.~\ref{numerics} we show results for the longitudinal and transverse resistivity as a function of the amplitude and the orientation of the in-plane magnetic field. Discussions and comments about the dependence of the calculations on the different parameters of the model are the subject of Appendix~\ref{parameters}. 
In Sec.~\ref{discussion} we analyse and highlight the competing effects of spin-orbit and magnetic fields on the electronic states at the Fermi level and, by consequence, on the scattering amplitudes
describing transitions between them. We isolate the different scattering mechanisms leading to the abundance of features observed in the angular-resolved magnetotransport. 
Sec.~\ref{conclusions} contains a final summary and outlook.  

\section{Anisotropic planar magnetotransport: experimental signatures}
\label{experimental_review}
To date there is a very broad literature \cite{brinkman2007, benshalom2009, benshalom2010, shalom2010, ariando2011, fete2012, joshua2012, joshua2013, annadi2013, fete2014, diez2015, yang2016} of experimental studies on the conducting interface of \ce{LAO}/\ce{STO} heterostructures -- grown along the $(001)$-crystalline axes -- in external magnetic field.
Here we restrict to low-temperature transport  (yet above the superconducting critical temperature $T_c \sim 300\ \si{\milli\kelvin}$) and magnetic field in the plane of the interface. 
A first striking observation is the qualitative change of transport properties that occurs when the system undergoes a Lifshitz transition \cite{joshua2012} by tuning the density of carriers via an applied gate voltage.
Joshua \emph{et al.} \cite{joshua2013} measured the in-plane angular dependence of the longitudinal ($\rho_{xx}$) and transverse ($\rho_{xy}$) resistivity  of a (back-gated) field-effect device by varying the angle between the current and the magnetic field within the plane of the interface.
At low gate-voltage $\rho_{xx}$ depend very weakly on the orientation $\phi_B$ of the magnetic field relative to the direction of the current, and the maximum and minimum resistivity are measured along the crystalline axes.
At high voltage the response of the system is extremely sensitive to the magnetic field: a large drop in $\rho_{xx}$ while increasing the field-strength occurs above a characteristic field $B_c$ of order of a few ${\rm T}$. The latter is shown to have a dependence on the gate-voltage $V_{\rm G}$, e.g. decreasing while increasing $V_{\rm G}$ and diverging while approaching the Lifshitz point from above. Moreover the magnetoresistance is strongly anisotropic and its angular modulation is considered as the signature of a change of some symmetry of the system. Additional peaks and dips points appears at intermediate angles. 
The percentage of anisotropy measured at $B=14\,\mathrm{T}$ is about $20 \%$ of the average resistivity. Along with that, the authors report an abrupt increase of the transverse resistivity $\rho_{xy}$ by increasing the field. Above $10\,{\rm T}$ $\rho_{xy}$ becomes comparable to $\rho_{xx}$ and characterized by a striking step-like angular modulation.  
The magnitude of $\rho_{xy}$ and its symmetry $(\rho_{xy} \simeq \rho_{yx})$ rule out any relevant contributions of the orbital field due either to minimal misalignment between the direction of $\bm B$ and the plane of the interface, or to the finite extension of the gas in the out-of-plane direction.
The \emph{crystalline} symmetry of the anisotropic response is revealed by the evolution of the direction of the principal axes of the resistivity tensor. While at low voltage (density) the principal axes follow the direction of $\bm B$, at high voltage and high magnetic field the principal axes are pinned to diagonal directions ($45^{\circ}, 135^{\circ}, 225^{\circ}, 315^\circ$): the directions where maximum and minimum resistivity are measured do not depend on the orientation of the magnetic field.

Similar behavior on different samples was previously reported by Ben Shalom \emph{et al.} \cite{benshalom2009} who also investigated the temperature dependence of the effect.
Sharp minima (maxima) of the longitudinal resistivity are measured when the magnetic field is perpendicular (parallel) to the current. The magnitude of the high-field anisotropy is consistent with the finding of Joshua \emph{et al.} \cite{joshua2013} and is suppressed on the same temperature-scale which governs the magnetoresistance.\cite{diez2015}

\section{Electronic structure and the Boltzmann equation with correlated disorder}
\label{theory_model}
\begin{figure*}[t]
\centering
\subfigure{
\includegraphics[width=0.35\textwidth]{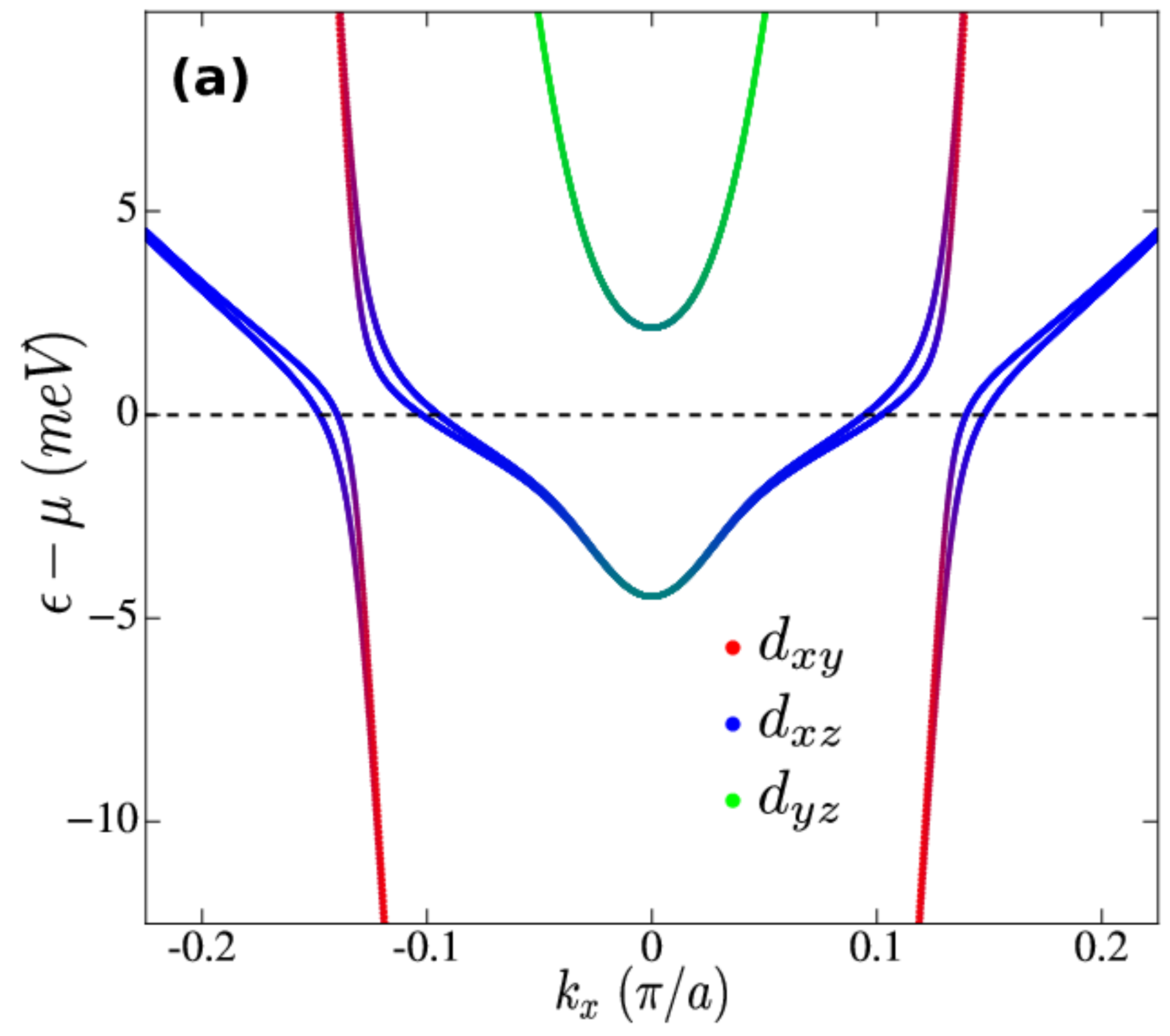}}
\subfigure[{\large{\bf SAM}} \ \ \ \ \ \ \ \ \ \ \ \ \ \ \ \ \ \ \ \ \ \ \large{\bf OAM}]{
\includegraphics[width=0.6\textwidth]{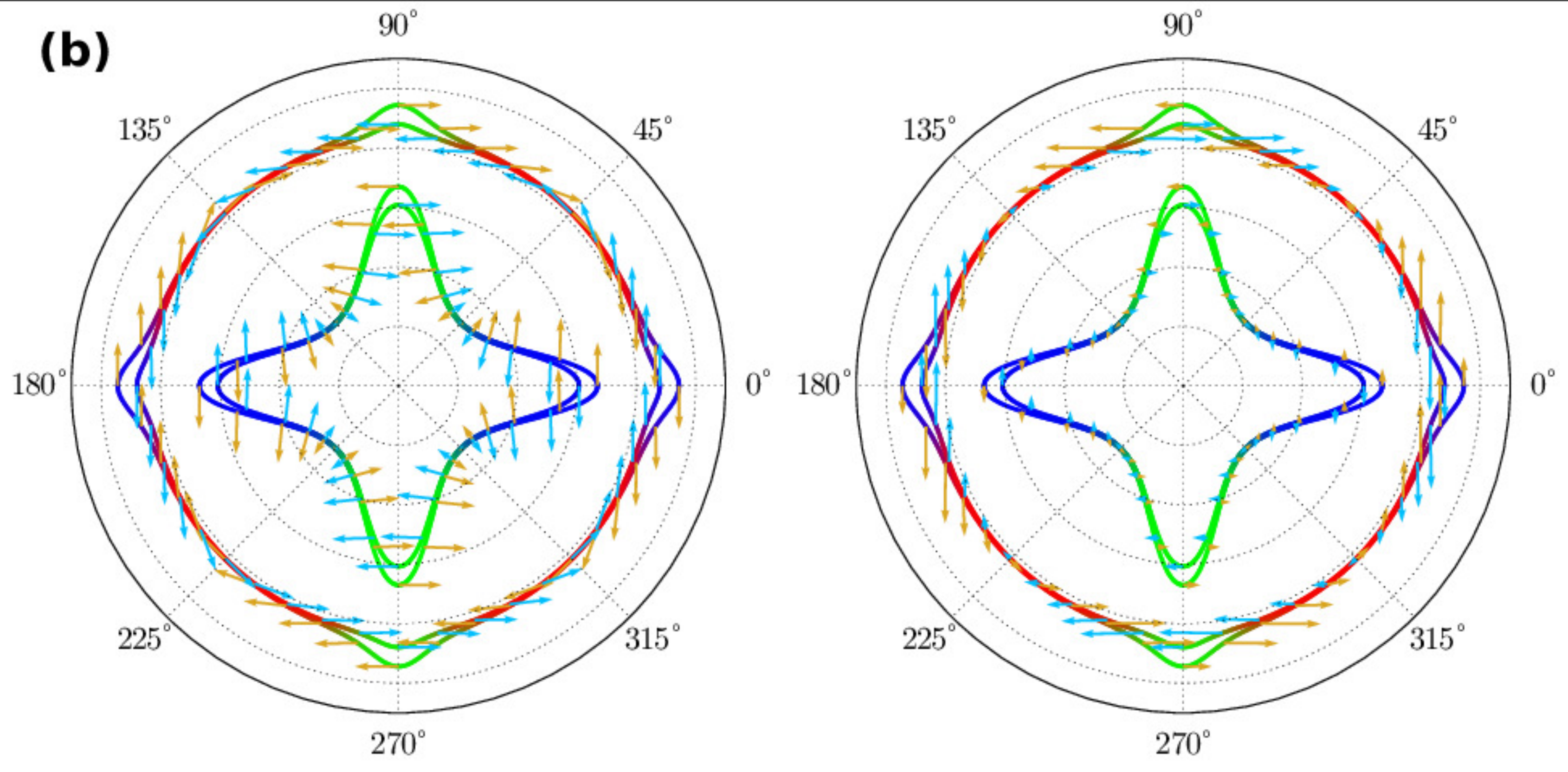}}
\caption{(a) Band dispersion along $k_x$ (at $k_y=0$) near the Lifshitz point. Colors distinguish the orbital character of the electronic states. (b) Equienergetic surfaces at $\epsilon=\mu$ (Fermi level) with on top textures of the the average spin ({\rm SAM}) and orbital ({\rm OAM}) angular momentum. Blue (yellow) arrows correspond to states in the outer (inner) band of each pair of subbands. The complete set of parameters used for generating the plots are listed in the first line of Table~\ref{param_table} (Appendix~\ref{parameters}).\label{BS_and_FS__H_0}}
\end{figure*}

The low-energy electronic structure is obtained from the single-particle Hamiltonian introduced by Ruhman \emph{et al.}\cite{ruhman2014} (in a three-orbital basis). The motions of the electrons is confined to the plane of the interface ($xy$ plane). 
(Recent ultra-high-field measurements \cite{yang2016} are consistent with a fully two-dimensional characterization of the mobile electron system.) 
In terms of creation (annihilation) operators $c_{{\bm k},l,\sigma}^\dagger$ ($c_{{\bm k},l,\sigma}$) of an electron with momentum ${\bm k}$ in the $l=(d_{xy}, d_{xz}, d_{yz})$ orbital of titanium atoms, the tight-binding Hamiltonian in presence of an external magnetic field 
\begin{align}
\mathcal{H} &= \sum_{\k,l, l', \sigma, \sigma'}c_{\k,l,\sigma}^\dagger \ 
H_{l\sigma,l'\sigma'}(\k) \ c_{\k,l',\sigma'} \nonumber \\
H &= H_{\rm L}+
H_{\rm SO}+H_{\rm Z}+H_{B}.
\label{ham}
\end{align}
is the sum of the kinetic term $H_{\rm L}$, the atomic spin-orbit coupling $H_{\rm SO}$, the inter-orbital coupling $H_{\rm Z}$ -- which allows electrons to hop from one metal site to another through intermediate oxygen atoms \cite{khalsa2013} due to inversion-symmetry-breaking at the interface -- and the Zeeman coupling
of the magnetic field with spin ({\rm SAM}) and orbital ({\rm OAM}) angular momentum. The matrix representations of these operators are shown in Appendix~\ref{model}. 
The energy spectrum near the Lifshitz point is plotted in Fig.~\ref{BS_and_FS__H_0}(a). The interface-confinement breaks the bulk degeneracy of $d_{xy}$ and $d_{xz, yz}$ states at $\k=0$, since the former are characterized by small hopping amplitude along the out-of-plane ($z$) direction and their energy at the $\Gamma$-point is lowered by an amount $\Delta_{\rm E}$.
At low density only $d_{xy}$ bands are populated and the effective spin-orbit interaction of ordinary Rashba-type, with coupling constant $\alpha_{\rm R} \sim \Delta_{\rm Z} \Delta_{\rm SO}/\Delta_{\rm E}$ \cite{kim2013} ($\Delta_{\rm Z}$ and $\Delta_{\rm SO}$ are the inversion-symmetry breaking and atomic spin-orbit parameter respectively, see Appendix~\ref{model}.)
A change in the topology of the Fermi surface, due to the onset of occupation of a new pair of bands,
occurs at a threshold density $n^*$.

The interplay of $H_{\rm SO}$ and $H_{\rm Z}$  produces strong orbital hybridization and spin-splitting for electronic states at points in the Brillouin zone where light and heavy bands would cross each other at $H_{\rm SO}=H_{\rm Z}=0$.
In the absence of magnetic field, analytical expressions for the effective Rashba-like coupling of the surface states of \ce{SrTiO3} and \ce{KTaO3} were derived by Kim \emph{et al.} \cite{PKim2014}. A similar derivation (valid near the $\Gamma$-point) was worked out by Zhou \emph{et al.} \cite{zhou2015}.
Here we resort to numerical diagonalization of the Hamiltonian including the Zeeman coupling $H_{\rm B}=\mu_{\rm B}(\bm L + g\bm S)\cdot \bm B / \hbar$ of the magnetic field $\bm B$ with the orbital ($\bm L$) and spin ($\bm S$) angular momentum.
As shown in Fig.~\ref{BS_and_FS__H_0}(b), at $T=0$ the Fermi level is characterized by two small surfaces, elongated along the symmetry axes of the crystal,  and two larger and less anisotropic ones. 
Importantly, the group velocity ${\bm v}_{\k, \nu} = \hbar^{-1} \partial \epsilon_{\k, \nu}/\partial {\k}$ is no longer parallel to the momentum for large sections of the Fermi surfaces.

We calculate the expectation-value of the spin and orbital angular-momentum operators on the eigenstates. At $B=0$, the $z$-component of both is quenched to zero because of time-reversal and $\pi$-rotation symmetry around the $z$-axis and will stay zero as long as the magnetic field has no component in the out-of-plane direction. 
Following the evolution of the average {\rm SAM} on the large Fermi surfaces in the top-right quadrant of the Brillouin zone ($\vartheta<90^{\circ}$), it is found to be parallel to the $y$-axis at small $\vartheta$ (small $k_y$), it suddenly undergoes a $90^{\circ}$-rotation in the vicinity of $\vartheta=45^{\circ}$ and finally aligns to the $x$-axis at $\vartheta > 45^{\circ}$. 
The magnitude of the {\rm OAM} is peaked near the hybridization gaps while it is very small on the remaining sectors of the Fermi surfaces.
Electronic spectrum and the spin-orbital structure at the Fermi level are consistent with the data reported by King \emph{et al.} \cite{king2014} for the surface states of \ce{SrTiO3}.

The eigenstates $|\psi_{\k,\nu}\rangle = |u_{\k,\nu}\rangle e^{i \k \cdot {\bm r}}$ and the eigenvalues  $\epsilon_{\k,\nu}$ of the Hamiltonian \ref{ham} enter the Boltzmann transport equation
\begin{equation}
  -e({\bm v}_{\k,\nu}\cdot {\bm E})\frac{\partial f_0}{\partial \epsilon_{\k,\nu}}=
\sum_{\k',\nu'}(g_{\k,\nu}-g_{\k',\nu'})q_{\k\nu,\k'\nu'}\delta(\epsilon_{\k,\nu}-\epsilon_{\k',\nu'}),\label{boltzmann}
\end{equation}
returning the shift in the electron distribution $g_{\k,\nu}$ due to the action of an accelerating electric field ${\bm E}$ and scattering by impurity centres. $f_0(\epsilon)$ is the equilibrium Fermi-Dirac distribution function and ${\bm v}_{\k,\nu} = \hbar^{-1} \partial \epsilon_{\k,\nu}/\partial \k$. 
Spatial correlations between different impurities can be introduced via a Gaussian potential
\begin{equation}
U(\bm r)=\sum_i U_i e^{-|\bm r-\bm r_i|^2/\xi^2}\,.  
\end{equation} 
where the amplitudes $U_i$ of the individual scatterers are randomly distributed with uniform probability in the symmetric range $[-\delta/2, \delta/2]$ and  $\xi$ is the characteristic decay-length of the two-point correlator (that is Gaussian as well).

At leading order in the Born approximation and averaging over the ensemble of impurity configurations,  the amplitude of elastic scattering from the initial state $|u_{{ \bm k}\nu}\rangle$ to the final state $|u_{{ \bm k'}\nu'}\rangle$ is 
\begin{equation}
 q_{\k\nu,\k'\nu'}=\tfrac{2}{3}\pi^3\hbar^{-1} \delta^2\xi^4 n_{\rm imp}\,e^{-\xi^2|\k-\k'|^2/2} |\langle u_{{\bm k}\nu}|u_{{ \bm k'}\nu'}\rangle|^2 \ ,
  \label{eq:qkkp}
\end{equation}
$n_{imp}$ being the density of impurities.

An alternative model was considered by Fu \emph{et al.} \cite{fu2016} who calculated the density-dependence of the resistivity in multi-subband accumulation layers (heterojunctions of polar and non polar perovskites such as \ce{LAO} and \ce{STO}) where electrons are scattered by the potential generated by surface roughness. In their model, spatial disorder-correlations decay exponentially. In momentum-space, the elastic scattering amplitude
\begin{equation}
q_{\k\nu,\k'\nu'} = \frac{2\pi^2 \delta^2 \xi^2}{\hbar}n_{imp}{(1+\xi^{2}{|\k -\k'|}^2)}^{-3/2} |\langle u_{\k \nu}|u_{\k'\nu'}\rangle|^2\ 
\label{eq:qkkp_exp}
\end{equation}
decays algebraically as a function of the momentum transferred to the impurity ($q_{\k\nu,\k'\nu'} \propto {|\k-\k'|}^{-3}$ at large $|\k-\k'|$).
The scattering with large momentum-transfer (and hence backscattering) is stronger for this model than for the Gaussian model of Eq.~\ref{eq:qkkp}.

We consider scatterers with no internal degrees of freedom (scalar impurities); nevertheless the (inter-band) scattering amplitudes with $\nu \neq \nu'$ are non-zero due to the off-diagonal hamiltonian elements in the orbital basis (Appendix~\ref{model}). 
Here we calculate the longitudinal and transverse magnetoresistivity for both models and find -- at fixed density -- close similarity in the outcomes, albeit the density-dependence of bare resistivities may be quite different in the two cases \cite{fu2016}.

\section{Numerical results}
\label{numerics}

At linear order in the electric field $\bm E$.  the electron distribution $g_{\k, \nu}$ is expressed in terms of the band- and momentum-dependent vector mean-free-path (\textit{vmfp}) $\bm\Lambda_{\k, \nu}$\cite{pikulin2011} as
\begin{equation}
g_{\k, \nu}=-e(\partial f_0/ \partial \epsilon_{\k, \nu}){\bm E} \cdot \bm\Lambda_{\k, \nu}.
\end{equation}
Eigenvalues $\epsilon_{\k, \nu}$ and eigenvectors $u_{\k, \nu}$ of the Hamiltonian \ref{ham} are calculated by numerical diagonalization. The components of the in-plane magnetic field are
\begin{equation}
{B}_x = {B}\cos\phi_{B} \,\,\,\,\,{B}_y = {B}\sin\phi_{B}
\end{equation} 
where we introduced the magnitude ${B}=|\bm B|$ and the angle $\phi_B$ measured counterclockwise from the $x$-axis.
The (mobile) electron density is constant (at any $B$) and, as a consequence, the chemical potential $\mu({B}, \phi_ {\rm B})\equiv\mu({B})$ is determined by demanding
\begin{equation}
n=\int_{\epsilon_0}^{\infty} d\epsilon\,f_0 \big(\epsilon, \mu(B), T \big)\, N(\epsilon)\ ,
\label{carrier_density}
\end{equation}
where $N(\epsilon)$ is the density of states at energy $\epsilon$ and $\epsilon_0$ the energy of the bottom of the lowest conduction band. 

\begin{figure*}[t]
\centering
\subfigure[{\bf \large Gaussian}]{
\includegraphics[width=0.45\textwidth]{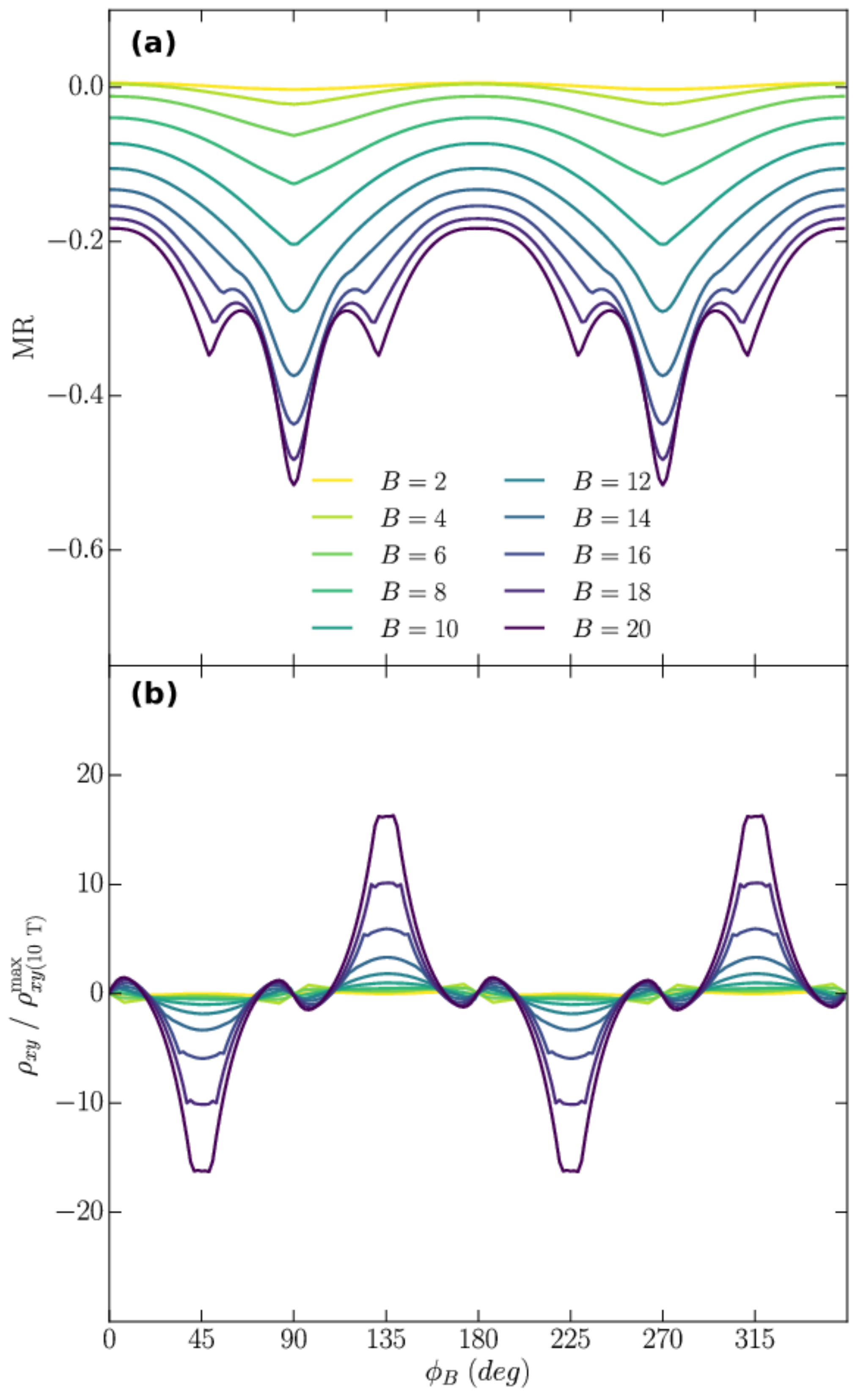}}
\subfigure[{\bf \large Exponential}]{
\includegraphics[width=0.435\textwidth]{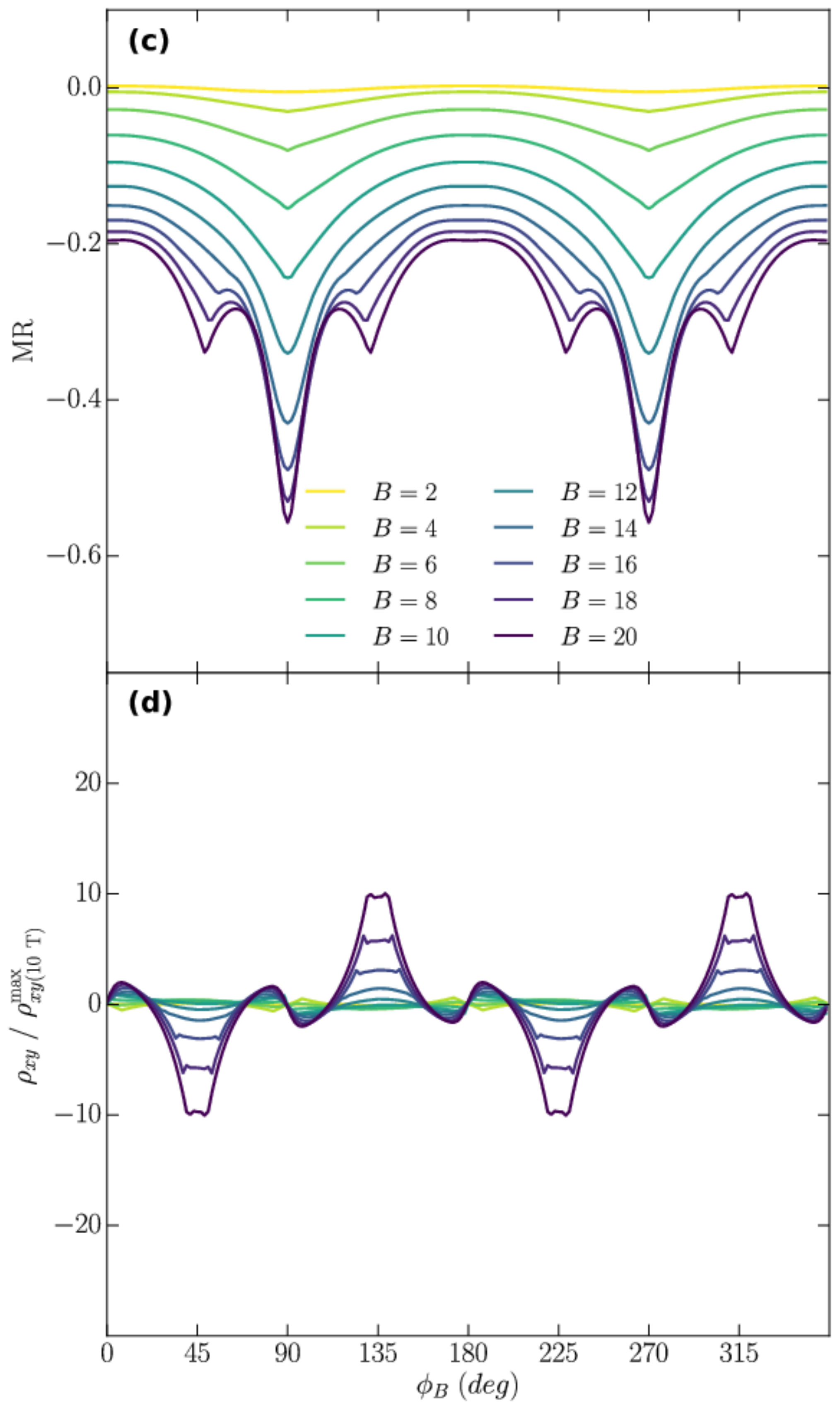}}
\caption{Angular modulations of {\rm \bf (a)}-{\rm \bf (c)} the longitudinal magnetoresistance ${\rm MR}$ and of
{\rm \bf (b)}-{\rm \bf (d)} the transverse resistivity $\rho_{xy}$ at several values of the $\bm B$-field in the range $2 - 20\ {\rm T}$. The temperature is $T=1\ \si\kelvin$ and the density $n = 2.2\times 10^{13}\,{\rm cm}^{-2}$ (chemical potential at half of the gap between $d_{xz}$ and $d_{yz}$ states at $\k=0$). 
In {\rm \bf (a)} and {\rm \bf (b)} scattering-amplitudes given by Eq.~\ref{eq:qkkp}.  In {\rm \bf (c)} and {\rm \bf (d)} scattering-amplitudes given by Eq.~\ref{eq:qkkp}.
Parameters listed in Table~\ref{param_table} (Appendix~\ref{parameters}).  \label{anisotropy_plot}}
\end{figure*}

From the distribution function the conductivity tensor $\bm\sigma$ follows as
\begin{equation}
  (\bm\sigma)_{ij}=e\sum_{{\bm k},\nu}(\bm{v}_{{\bm k},\nu})_i\frac{\partial g_{{\bm k},\nu}}{\partial E_j}.
\label{sigma_ij}
\end{equation}
By inverting $\bm\sigma$ we finally calculate longitudinal and transverse resistivity
\begin{equation}
\rho_{xx} = \frac{\sigma_{yy}}{\sigma_{xx}\sigma_{yy}-\sigma_{xy}^2}\,\, ,\,\,\rho_{xy} = -\frac{\sigma_{xy}}{\sigma_{xx}\sigma_{yy}-\sigma_{xy}^2}.
\end{equation}
We show below results of calculations of the magnetoresistivities at temperature $T=1\ \si\kelvin$ and carrier-density $n = 2.2\times 10^{13}\,{\rm cm}^{-2}$, where the Fermi level cross the second pair of bands and is approximately at the middle of the gap (at $\k=0$) between $d_{xz}$ and $d_{yz}$ states (dashed line in Fig.~\ref{BS_and_FS__H_0} (a)). Two other bands remain a few ${\rm meV}\,(\approx \Delta_{\rm SO}/2)$ above $\mu$ and do not play any role here since we only consider elastic scattering. (The gap is larger than the thermal broadening  of the Fermi-Dirac distribution at low-temperature). 
In Fig.~\ref{anisotropy_plot} the magnetoresistance ${\rm MR}=\rho_{xx} (B)/\rho_{xx}(0) - 1 $ and the rescaled transverse resistivity $\rho_{xy} / \rho^{max}_{xy\ (10\,{\rm T})}$ are shown as a function of the angle $\phi_{B}$ and $B$ between $2$ and $20\ {\rm T}$, for the two scattering models \ref{eq:qkkp} and \ref{eq:qkkp_exp}. $\rho_{xy}$ is rescaled by its maximum value over the angular range ($\rho_{xy}^{max} \approx \rho_{xy}(\phi_B=45^{\circ}$)) at $10\, {\rm T}$ in order to get rid of the dependence of the calculations on the parameters $n_{imp}$ and $\delta$. 

In the range $4-10\,{\rm T}$ (at lower fields the effects are moderate) the angular modulation of the longitudinal ${\rm MR}$ has cusp-like dips at $\phi_B=90^\circ,270^\circ$ (magnetic field perpendicular to the current) and rounded maxima at $\phi_B=0^\circ, 180^\circ$ (magnetic field aligned to the current). The magnitude of the negative ${\rm MR}$ and the anisotropy progressively increase with the field-strength. The transverse resistivity has a sinusoidal modulation with maxima and minima shifted by $45^{\circ}$ with respect to $\rho_{xx}$. However $\rho_{xy}$ is about a factor $100$ smaller than $\rho_{xx}$.
Above $ 10\,{\rm T}$ the angular {\rm MR} develops additional maxima and minima near diagonal orientations $(\phi_{B} = 45^{\circ},135^{\circ},225^{\circ},315^\circ)$ that unlike the main extremal points -- fixed at multiples of $90^\circ$ -- do not only move up and down but also shift in angular position as the field is progressively increased. In the same field-range where these additional features characterize the ${\rm MR}$,  $\rho_{xy}$ increases by (more than) one order of magnitude. Another striking feature is the change in the angular modulation of the transverse resistivity that substantially deviates from the sinusoidal low-field behaviour. 

\begin{figure*}
\begin{center}
\subfigure[{\bf \large Gaussian}]{
\includegraphics[width=0.45\textwidth]{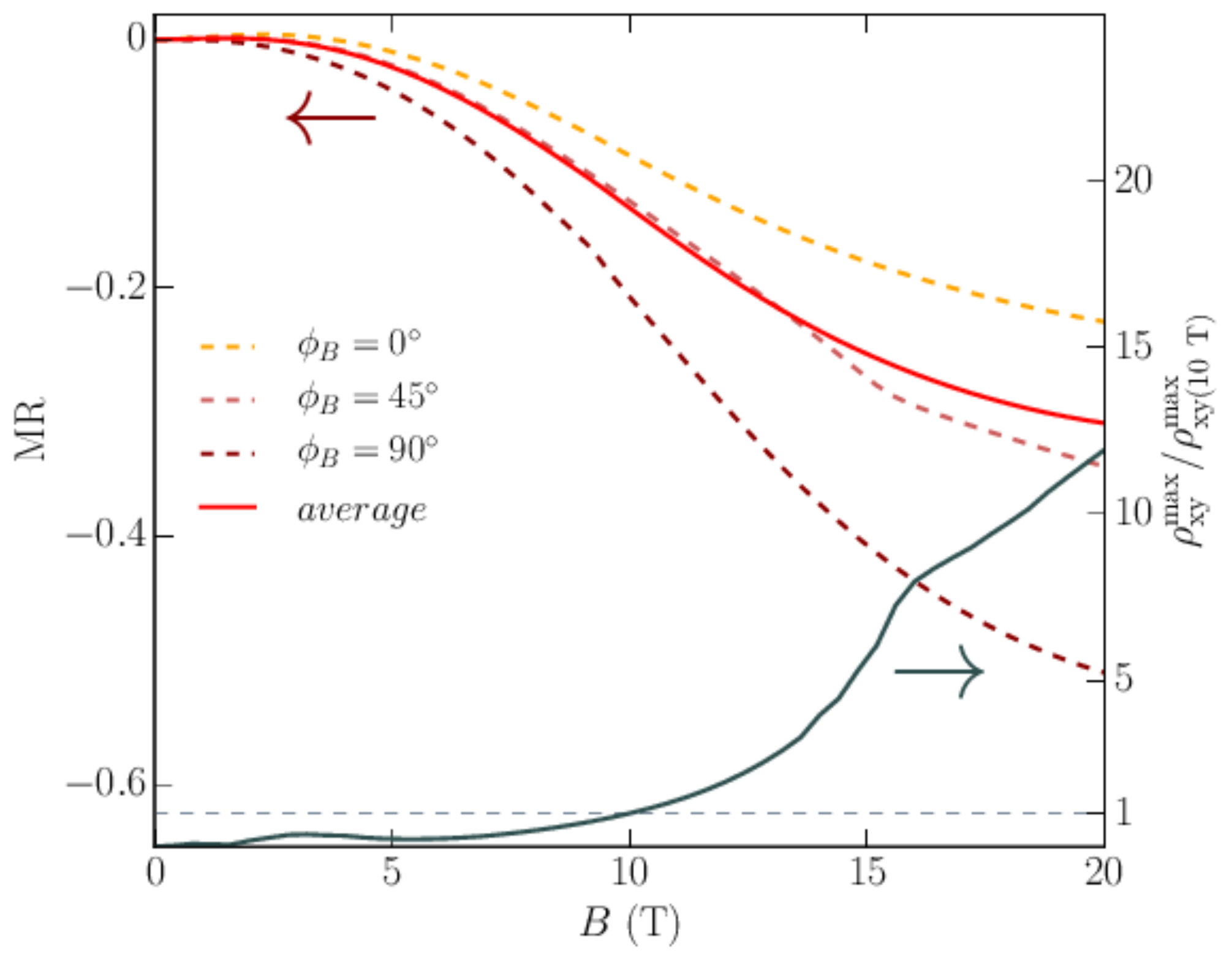}}
\hspace{3mm}
\subfigure[{\bf \large Exponential}]{
\includegraphics[width=0.435\textwidth]{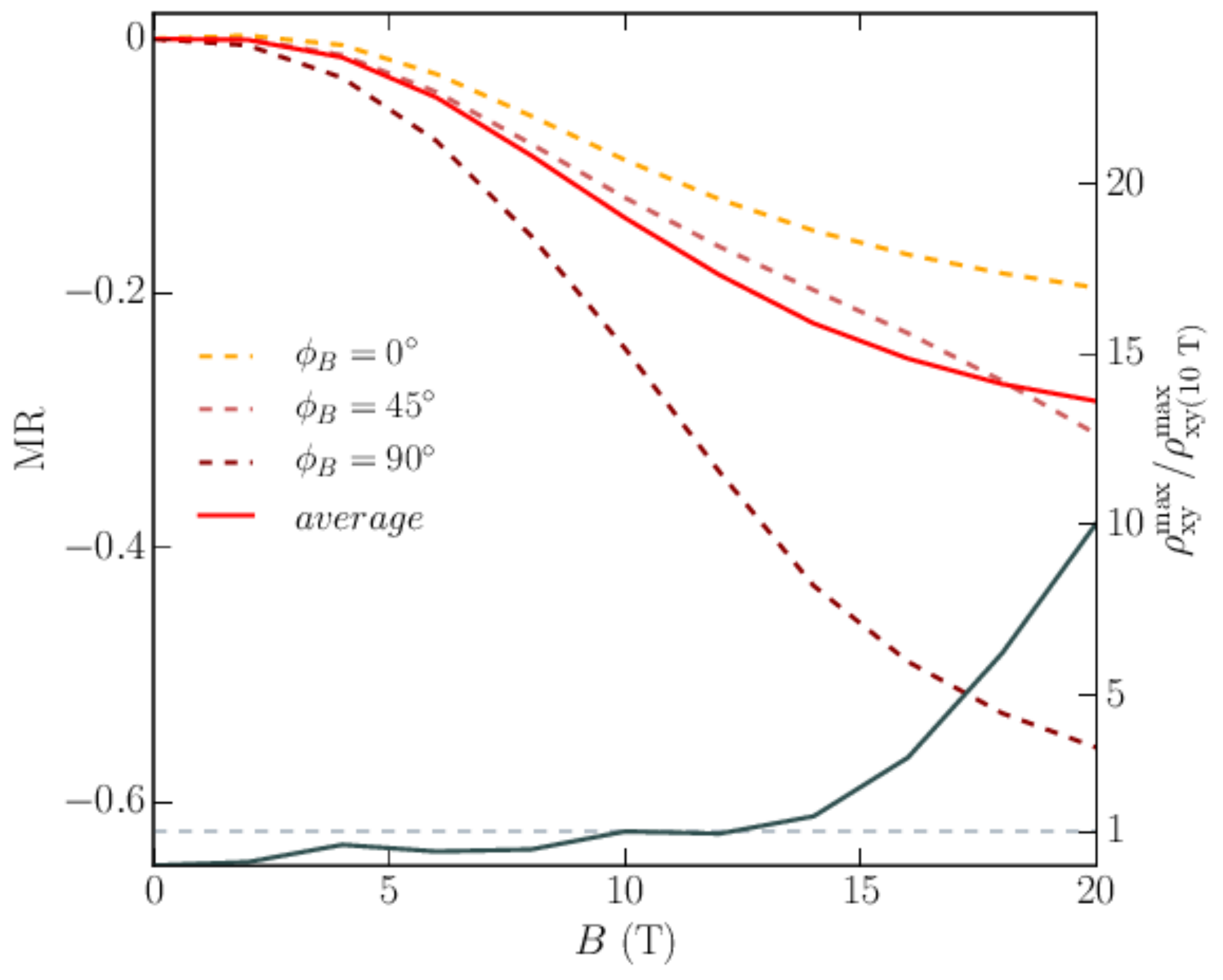}}
\caption{Longitudinal magnetoresistance at $T = 1 \ \si\kelvin$ for different orientations $\phi_B$ (dashed lines) and averaged over all the angles (red solid line). Numerical values on the left axis. Angular-maximum of $\rho_{xy}$ as a function of $B$ rescaled by its value at $B=10 \ {\rm T}$ (solid grey line) with numerical values on the right scale. A comparison between calculations for scattering models \ref{eq:qkkp} and \ref{eq:qkkp_exp} is shown.
\label{MR_phiB}}
\end{center}
\end{figure*}

The $B$-dependence of the magnetoresistance at different angles $\phi_B$ is shown in Fig.~\ref{MR_phiB}. Amplitude, shape and the field-scale of the ${\rm MR}$, all change with $\phi_B$. Moving from a configuration  with magnetic field parallel to the direction of the current ($\phi_B = 0^\circ$) towards the opposite configuration ($\phi_B=90^\circ$), the magnitude of ${\rm MR}$ grows by a factor $\sim 3$ at $20\,{\rm T}$. Moreover, the field-scale where the slope of ${\rm MR}$ becomes negative -- and large in magnitude -- decreases by moving the field away from the direction of the current.
The results are consistent with previous calculations\cite{diez2015} of the ${\rm MR}$ at $\phi_{B}=90^{\circ}$, albeit here calculated in a different {\rm SOC}-regime (see Appendix~\ref{parameters}).
The angular-maximum of the transverse resistivity $\rho_{xy}$ exhibits a strong enhancement around starting at $\sim 10\,{\rm T}$ for the Gaussian-model (slightly higher field for the exponential model). 
The slope of $\rho_{xy}(B)$ softens at the very high fields, where the magnetoresistance at $\phi_B=45^{\circ}$ shows the onset of saturation in the case of Gaussian disorder potential. For the other model, instead, the curvature of the ${\rm MR} (\phi_B=45^{\circ}$) is still negative at $B=20\, {\rm T}$ (no saturation in this field-range) that also produces a non-saturating $\rho_{xy}$.

\section{Discussion}
\label{discussion}
At $T=0$ all of the conductivity of a metal is effectively carried only by electrons at the Fermi level. In favour of a clearer discussion, hereafter we neglect the effects of low finite temperature -- these are crucial quantitatively, but do not alter the underlying mechanism. Therefore we can consider only electronic states at the Fermi level, characterized by $\epsilon_{\k, \nu} = \mu(n, B)$ (where $\mu$ is determined at any $B$ self-consistently according to Eq.~\ref{carrier_density}).

At low-density where only the lowest $d_{xy}$ states are populated, the spin-orbit coupling is weak and produces two nearly degenerate chiral bands with $k$-linear Rashba splitting \cite{kim2013}.  In this limit, anisotropic scattering does not produce anisotropic magnetoresistance even in the presence of spin-selective scatterering.\cite{trushin2009} Results of low-density calculations are shown in Fig.~\ref{lower_densities} in Appendix~\ref{parameters}.

In Sec.~\ref{theory_model} it was highlighted that above the Lifshitz point -- and particularly when the chemical potential is approximately at the middle of the gap between $d_{xz}$ and $d_{yz}$ states -- spin-orbit coupling is sharply enhanced around particular directions in the Brillouin zone (hybridization gaps).
For both models of disorder disorder backscattering in the outer bands is suppressed at $\xi \gg 1/k_F^{out}$ ($k_F^{out}$ is the average Fermi-momentum in the outer bands which are almost fully isotropic). Scattering processes with small transfer of momentum $|\k-\k'|$ coincide with low-angle processes. (This is not the case for the highly anisotropic inner bands.)
Because of the small (intra-band) backscattering amplitudes for electrons in the outermost bands, effective current-relaxation is achieved through forward (inter-band) scattering to the innermost bands. The latter have low mobilities due to the small velocities and large intra-band scattering rates (small $|\k-\k'|$). 
\begin{figure*}[t]
\centering
\subfigure[{\large $\,\,\,\,\,\,\,\,\,\,\,\,\,\, B=0\,{\rm T}\,\,\,\,\,\,\,\,\,\,\,\,\,\,\,\,\,\,\,\,\,\,\,\,\,\,\,\,\,\,\,\,\,\,\,\,\,\,\,\,\,\,\,\,\,\, \,\,\,\,\, \,\,\,\,\,\,\,\,\,\,\,\,\,\,\,\,\,\,\,\,\,\,\,\,\,\,B=20\,{\rm T}\,\,\,\,\,\,\,\,\,\,\,\, \,\,\,\,\,\,\,\,\,\,\,\,\,\phi_B=90^{\circ}$}]{
\includegraphics[width=0.8\textwidth]{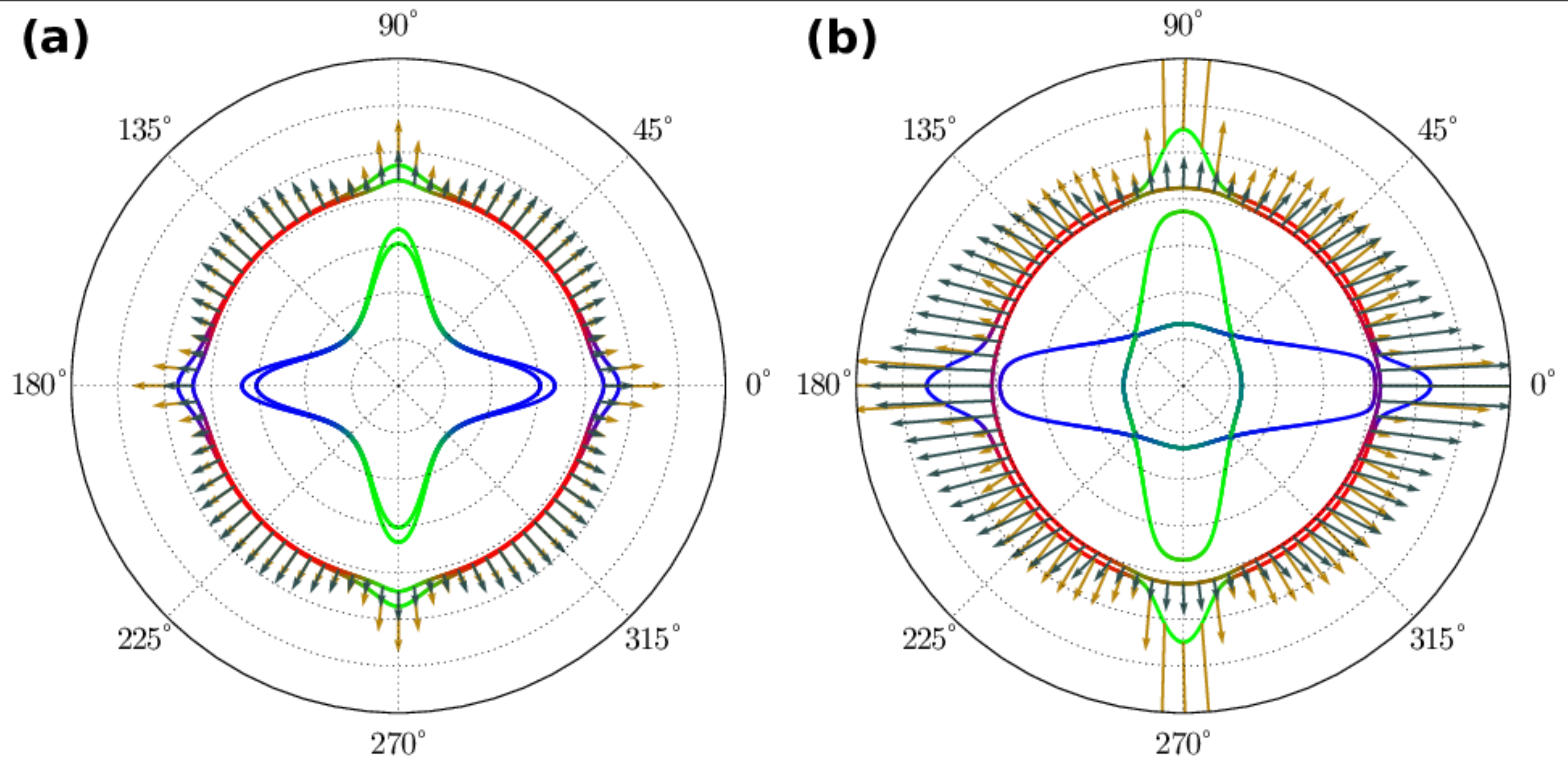}}
\subfigure[{\large $\,\,\,\,\,\,\,\,\,\,\,B=10\,{\rm T}\,\,\,\,\,\,\,\,\,\,\,\, \,\,\,\,\,\,\,\,\,\,\,\,\,\phi_B=45^{\circ}\,\,\,\,\,\,\,\,\, \,\,\,\,\, \,\,\,\,\,\,\,\,\,\,\,B=20\,{\rm T}\,\,\,\,\,\,\,\,\,\,\,\, \,\,\,\,\,\,\,\,\,\,\,\,\,\phi_B=45^{\circ}$}]{
\includegraphics[width=0.8\textwidth]{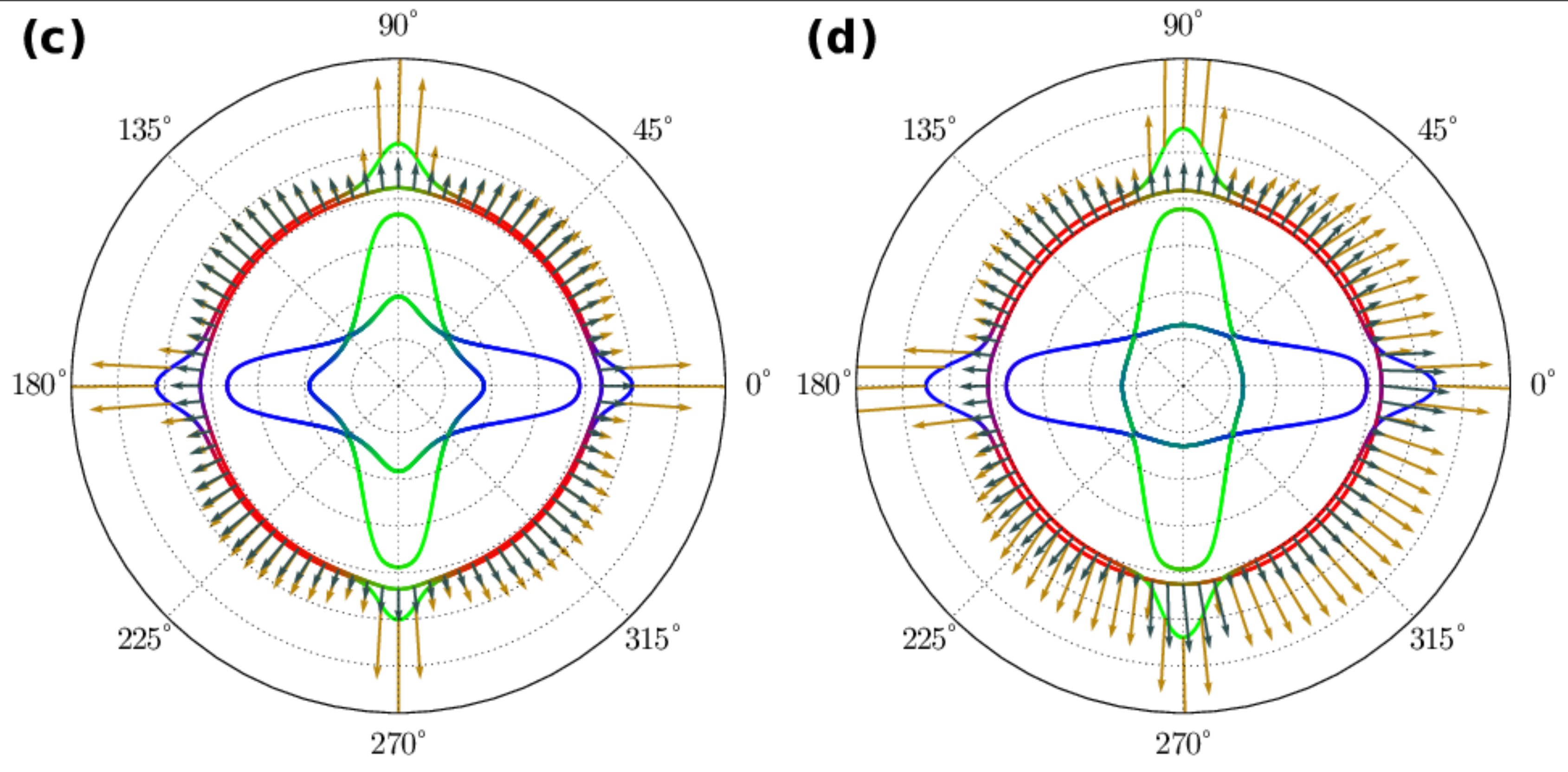}}
\caption{Fermi surfaces at $n = 2.2\times 10^{13}\,{\rm cm}^{-2}$ for different configurations of the in-plane magnetic field: {\rm \bf (a)} $B=0\,{\rm T}$, {\rm \bf (b)} $B=20\,{\rm T}\ \ \phi_{B} = 90^{\circ}$,
{\rm \bf (c)} $B=10\,{\rm T}\ \ \phi_{B} = 45^{\circ}$, {\rm \bf (d)} $B=20\,{\rm T}\ \ \phi_{B} = 45^{\circ}$. Magnitude and direction of the vector mean-free-path $\bm\Lambda_{\k}$ as a function of the momentum $\k$ on the outermost bands (which support all of the total conductivity at any fields) are represented by arrows - colors distinguish the two bands of each pair.
The longitudinal conductivity $\sigma_{xx}$ is proportional to the average $x$-component of the vector mean-free-path. (At $\phi_B=0,\,90^{\circ}$ $\rho_{xx}=\sigma_{xx}^{-1}$.)
The modulation of the \textit{vmfp} in {\rm \bf (c)} is mirror-symmetric with respect to the crystalline axes; hence $\sigma_{xy} (\rho_{xy}) \approx 0$ because states with opposite velocities compensate each other with equal weights $\Lambda^x$. Instead in {\rm \bf (d)} the texture of the \emph{vmfp} clearly violates the symmetry resulting in a finite and large $\sigma_{xy}$.\label{fermi_surfs_with_lambdas}}
\end{figure*}

With the electric field in the $x$ direction, the trend of the longitudinal (transverse) resistivity as a function of ${\bm B}$ follows the field-dependence of the non-equilibrium distribution $g_{\k, \nu}$ on sections of the large (outermost) Fermi surfaces where the component of the band-velocity ${\bm v}^x_{\k, \nu}$ (${\bm v}^y_{\k, \nu}$) is large. As evident from Fig.~\ref{anisotropy_plot} and Fig.~\ref{MR_phiB}, the field-scale where negative magnetoresistance sets on is angle-dependent, while the largest ${\rm MR}$ always occurs at $\phi_B=90^{\circ}, 270^{\circ}$.

At $B = 0$ the spin-ordering on neighbouring Fermi surfaces around the avoided crossings of $d_{xy}$ and $d_{xz,yz}$ states in the top-right quadrant of the Brillouin zone is
$| \downarrow \rangle | \uparrow \rangle | \uparrow  \rangle | \downarrow \rangle $ ({\rm SAM} in Fig.~\ref{BS_and_FS__H_0}(b)). By switching up the magnetic field, the ordering is reversed for the inner bands, precisely at points in the Brillouin zone where the component of the magnetic field on the (local) axis of spin-orbit field $\bm\Omega^{\rm SO}_{\k}=\Omega^{\rm SO}_{\k}\hat{\bm n}_{\k}$ fulfils the condition 
\begin{equation}
\bm B \cdot \hat{\bm n}_{\k} \gtrsim \Omega^{\rm SO}_{\k}\ .
\label{B_so}
\end{equation}
This leads to $| \downarrow \rangle  | \uparrow \rangle | \downarrow \rangle | \uparrow \rangle$ spin-ordering near the hybridization gaps and to reduction of the total amount of inter-band scattering between pairs of states $(\k, \nu)$ and $( \k', \nu')$ which have minimal inter-band distances $|\k-\k'|$ and parallel spins at $B=0$.  

The suppression of the scattering on a scale set by Eq.~\ref{B_so} is most effective for enhancing (reducing) the longitudinal conductivity (resistivity) at angles $\phi_B$ such that at points in the Brillouin zone where ${\bm B}$ is perfectly aligned to $\bm\Omega_{\k}^{\rm SO}$ the component $v^x_{\k,\nu}$ of the band-velocity is maximum ($\phi_B=90^{\circ},270^{\circ}$).

At higher fields (above $10\ {\rm T}$) the \emph{intra-band} scattering between pairs of states in the two outermost bands becomes relevant. 
In Fig.~\ref{fermi_surfs_with_lambdas} the calculated vectors mean-free-path $\bm\Lambda_{\k,\nu}$ are plotted on top of the corresponding Fermi surfaces for different configurations of the magnetic field.
In particular, when the field is oriented far from the crystalline axes the spin-orbital splitting in the outer bands is selectively modulated, depending on the relative orientation of $\bm B$ and $\bm\Omega_{\k}^{\rm SO}$. 

With reference to the plot in {\rm \bf (d)}, it is important to notice that the two outer bands cross each other at an angle $\vartheta=\overline{\vartheta}\approx 10^\circ$, while they are split along the complementary direction $(90^{\circ} - \overline{\vartheta})$ (top-right quadrant). 
The amplitudes of scattering between $d_{xy}$ states (red section) to hybrid $d_{xy}/d_{xz}$ states at $\vartheta\approx \overline{\vartheta}$ are enhanced by the magnetic field than compared to the amplitudes of scattering to states at $\vartheta\approx (90^{\circ}-\overline{\vartheta})$. 

To better illustrate the consequences of this we formally express the solution of the Boltzmann equation~\eqref{boltzmann} in the recursive form

\begin{widetext}
\begin{align}
\bm\Lambda_{\k, \nu} = {\bm v}_{\k, \nu} \tau_{\k, \nu} \ &+ \sum_{\k'\neq \k} q_{\k\nu, \k'\nu'}
\Big \lbrace {\bm v}_{\k', \nu'} \tau_{\k', \nu'}
\ + \sum_{\k''\neq \k, \k'} q_{\k'\nu',\k''\nu''} \Big\lbrace {\bm v}_{\k'', \nu''} \tau_{\k'', \nu''} \nonumber\\
&+ \sum_{\k'''\neq \k, \k', \k''} q_{\k''\nu'', \k'''\nu'''}\,{\bm v}_{\k''', \nu'''}+ \cdots \Big\rbrace \Big\rbrace,
\label{full_lambda}
\end{align}
\end{widetext}
where $\tau_{\k, \nu}$ is the bare band- and momentum-dependent relaxation-time
\begin{equation}
\tau_{\k, \nu} = \sum_{\k', \nu'} q_{\k,\nu, \k'\nu'}.
\end{equation}
When the low-angle scattering is anisotropic, scattering-in corrections to $\bm\Lambda^{\rm RTA}_{\k,\nu}=\tau_{\k,\nu}\bm v_{\k,\nu}$ calculated in relaxation-time approximation ({\rm RTA})\cite{ziman1972,ziman1961} substantially affect magnitude and direction of the vector mean-free-path, that gets tilted (away from the direction of the velocity $\bm v_{\k,\nu}$) towards the direction of enhanced scattering ($\overline{\vartheta}$). The anisotropy of the scattering effectively acts as a force in momentum-space driving a shift of the electron distribution around the Fermi surface, similarly to the action of the Lorentz force in real-space when a magnetic field is applied perpendicular to the plane.
This mechanism is sometimes referred to as \emph{effective Lorentz force} ({\rm ELF}) and is known to be responsible for transport anomalies in multiband systems \cite{varma2001, breitkreiz2014}. 
However it is usually investigated in addition to an actual orbital field.
By gradually varying $B$ and $\phi_B$, magnitude and direction of the {\rm ELF} at a point $(\k, \nu)$ change non monotonically. In other words, the scattering amplitudes are non-monotonic functions of the magnetic field and the direction of the tilting of $\bm\Lambda_{\k,\nu}$ undergoes multiple reversals when the magnetic field changes in magnitude and/or direction. 

Therefore, 
the electron distribution
\begin{equation}
g_{\k, \nu} \propto \bm\Lambda_{\k, \nu} \cdot {\bm E}
\label{g_propto_lambda_dot_E}
\end{equation}
may increase or decrease depending on whether the tilting points to the $x$ or $y$-axis, respectively.
The appearance of secondary maxima and minima of the {\rm MR} at intermediate orientations of the magnetic field, and the shift in their angular positions as a function of $B$, follow by the dynamics of the {\rm ELF}-texture around the Fermi surfaces.

Finally, let us discuss what are the effects on $\rho_{xy}$ which gets no contribution from the inter-band scattering. It is uniquely the intra-band scattering that is responsible for the planar Hall-effect (transverse resistivity in presence only of in-plane magnetic field).
We need to look at the total contribution of states with opposite velocities in the direction orthogonal to the electric field and understand why an imbalance between their occupations is generated by the magnetic-field. With reference to states in the bottom-right quadrant ($270^{\circ}<\vartheta<360^{\circ}$), 
the scattering-in corrections are large and isotropic; hence they bring a total contribution to the magnitude but, importantly, no contribution to the direction of $\bm\Lambda$. According to Eq.~\ref{full_lambda}, these states have $\bm\Lambda_{\k,\nu} \propto \bm v_{\k,\nu}$  (as it would be according to {\rm RTA}) but the magnitudes of the vectors are much larger than the corresponding {\rm RTA} results. 
Similarly, states in the top-left quadrant of the Brillouin zone also have $\bm\Lambda_{\k,\nu} \propto \bm v_{\k,\nu}$. However the scattering-in corrections are smaller for such states because the magnetic field does not suppress -- but rather enhances -- the spin-orbital splitting at any point in the quadrant.  
From Eq.~\ref{g_propto_lambda_dot_E} (with $\bm E \parallel \hat{x}$) we conclude that an imbalance is produced between the occupation of electronic states with velocity $-v_{\k}^y$ (bottom-right) and the occupation of states with velocity $v_{\k}^y$ (upper-left).
that leads to a large $\rho_{xy}$ ($~\sim 0.1\cdot\rho_{xx}$ at $B=20\,{\rm T}$ and parameters in Table~\ref{param_table}). 

At constant $B$, the angular modulation of $\rho_{xy}$ sharply ramp up when the angle $\phi_B$ takes a value such that at isolated points in the Brillouin zone the spin-orbital splitting is totally suppressed by the Zeeman field. Once this occurs, the low-angle scattering suddenly becomes strongly anisotropic and then remains stable until the field is rotated far enough to let the spin-orbital splitting open again. This results in a flattening of the peaks of $\rho_{xy}$ more or less pronounced in all the plots shown in this paper (Figs.~\ref{anisotropy_plot},\ref{g_negative},\ref{more_xi},\ref{lower_densities}) 

\section{Conclusions}
\label{conclusions}
In this work, we investigated magnetotransport at the conducting interface of heterostructures formed by transition-metal oxides \ce{LaAlO3} and \ce{SrTiO3} and, in particular, the dependence of the resistivity tensor on the magnitude and orientation of an in-plane magnetic field.

Effects of many-body interactions (e.g. electron-electron \cite{maniv2015, tolsma2016} or magnetic couplings \cite{joshua2013, ruhman2014})are neglected, while spin-orbit coupling is treated with a microscopic model of the low-energy conduction states including six electronic bands. 
The Boltzmann equation (\ref{boltzmann}) for electrons scattered by correlated impurities -- with correlations decaying on a characteristic length scale $\xi$ -- is solved numerically as a function of the external field. Our main finding is a crossover from the low-field regime of weak anisotropy to the high-field regime of strong anisotropy which results from important changes in the electronic structure at the Fermi level when the carrier-density is tuned above the Lifshitz point.\cite{ruhman2014} 
However, we remark that is not simply the onset of the occupation of anisotropic bands to determine a change of the magnetotransport properties, but rather the selective modulation of the impurity-scattering connecting pairs of states at the Fermi level. 
In presence of magnetic field, the spin-orbital structure of the Bloch states is locally modified depending on the relative orientation of spin-orbit and magnetic fields. In particular, the effect is enhanced in the vicinity of (avoided) band crossings when the component of the magnetic field on the local (band- and momentum-dependent) axis of the spin-orbit field is comparable to the magnitude of the latter. Scattering amplitudes \emph{to} and \emph{from} these states are then extremely sensitive to the magnitude and direction of the magnetic field.

Our results are in good qualitative agreement with experiments,\cite{joshua2013} although some features of the experimental data remain not fully captured by our simple model.


The physics of complex oxide interfaces is interesting and promising for the development of electronic devices at the nanoscopic length scale.
Our work highlights the richness of electronic transport characterizing the system due to the unconventional structure of its spin-orbit coupling. It hopefully provides a further step towards a better understanding of the role of spin-orbit interaction in the conducting properties of oxide interfaces. 

We have benefited from discussions with C. W. J. Beenakker, A. R.
Akhmerov, A. D. Caviglia, A. M. R. V. L. Monteiro, M. Breitkreiz and E. Cobanera.
This research was supported by the Foundation for Fundamental Research on Matter (FOM), the Netherlands Organization for Scientific Research (NWO/OCW), and
an ERC Synergy Grant.
\vspace{5cm}
\appendix
\section{Single-particle Hamiltonian}
\label{model}
\begin{figure*}[t]
\subfigure{
\includegraphics[width=0.8\textwidth]{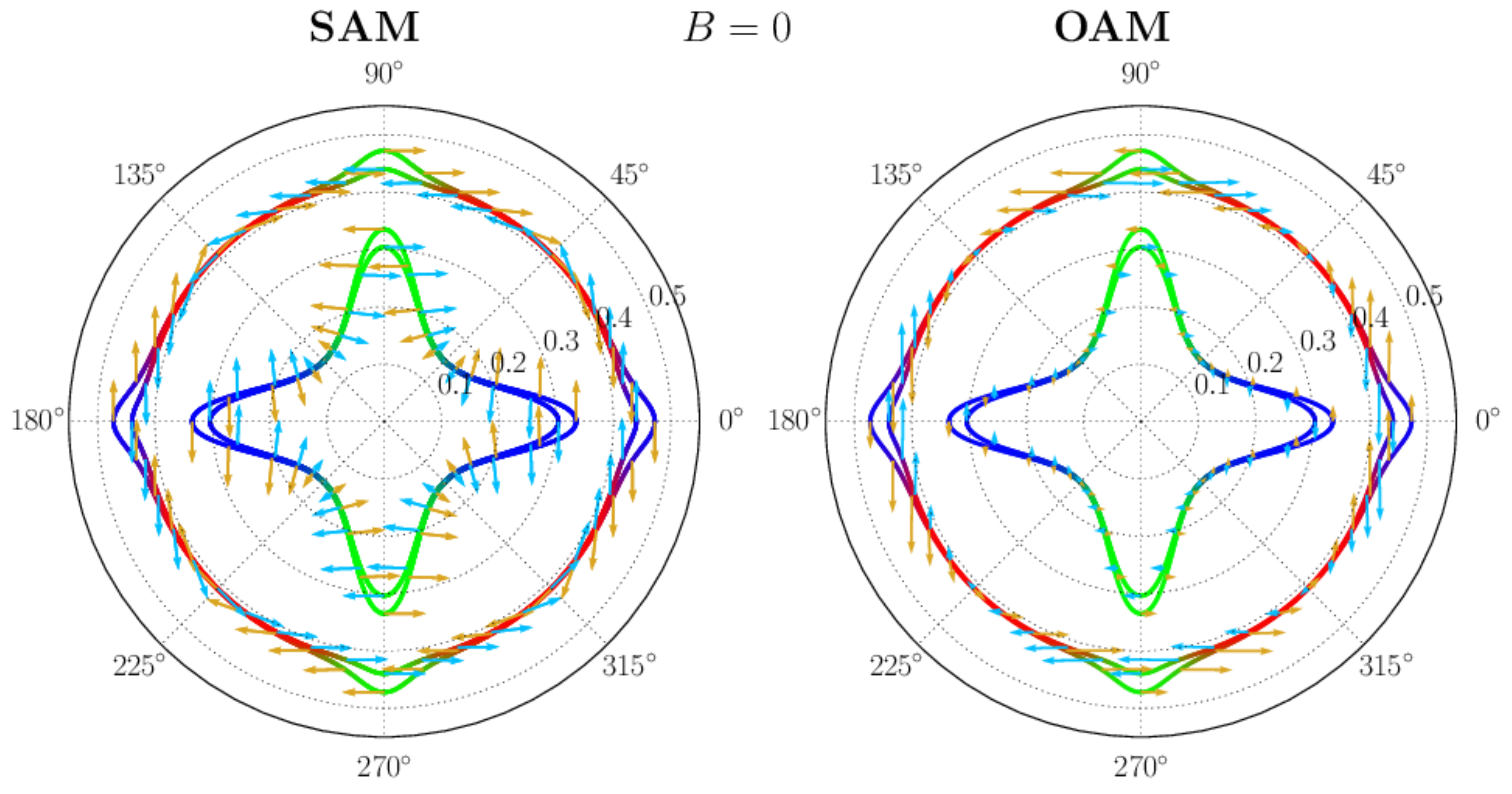}}
\subfigure{
\includegraphics[width=0.8\textwidth]{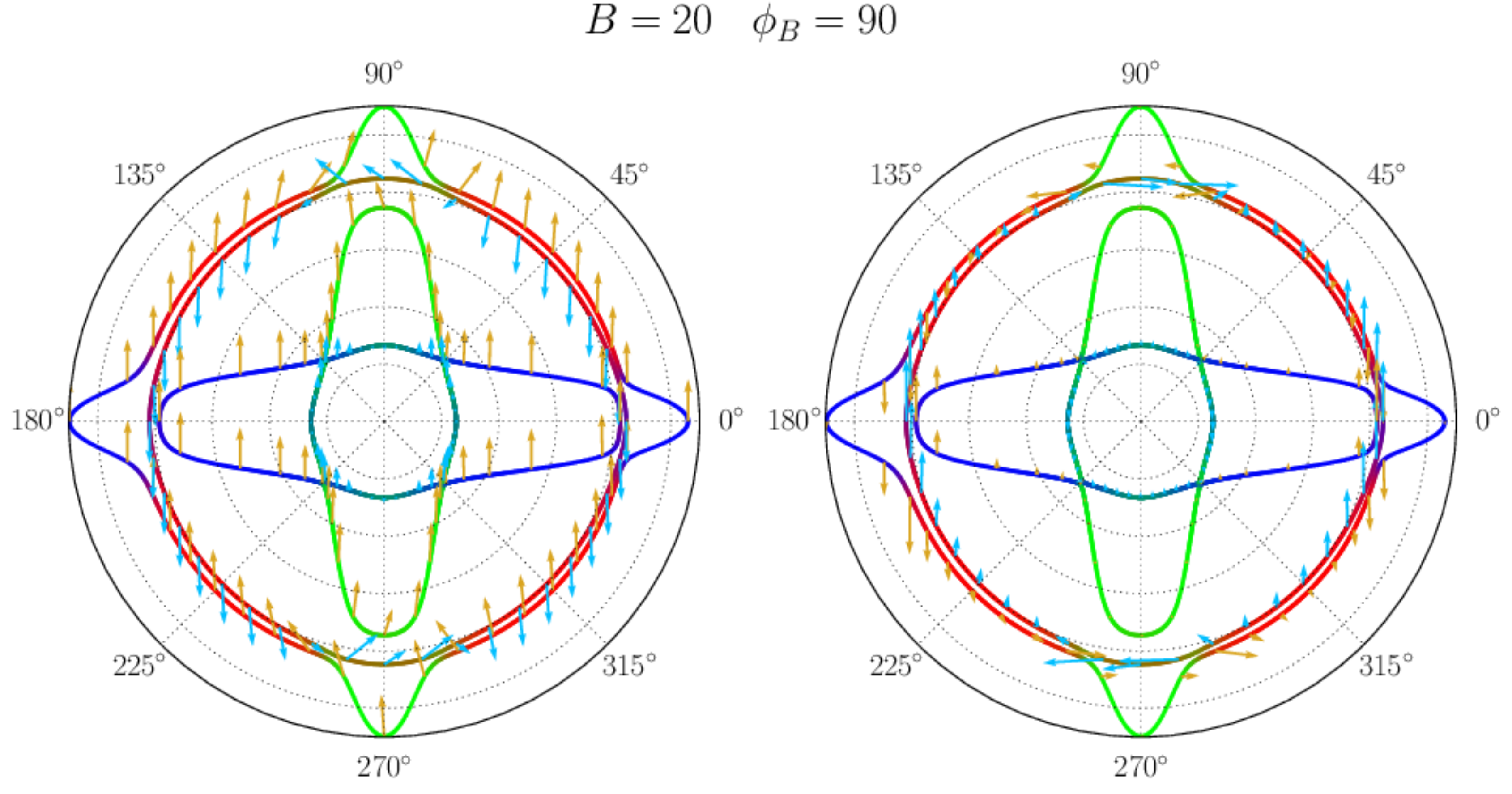}}
\subfigure{
\includegraphics[width=0.8\textwidth]{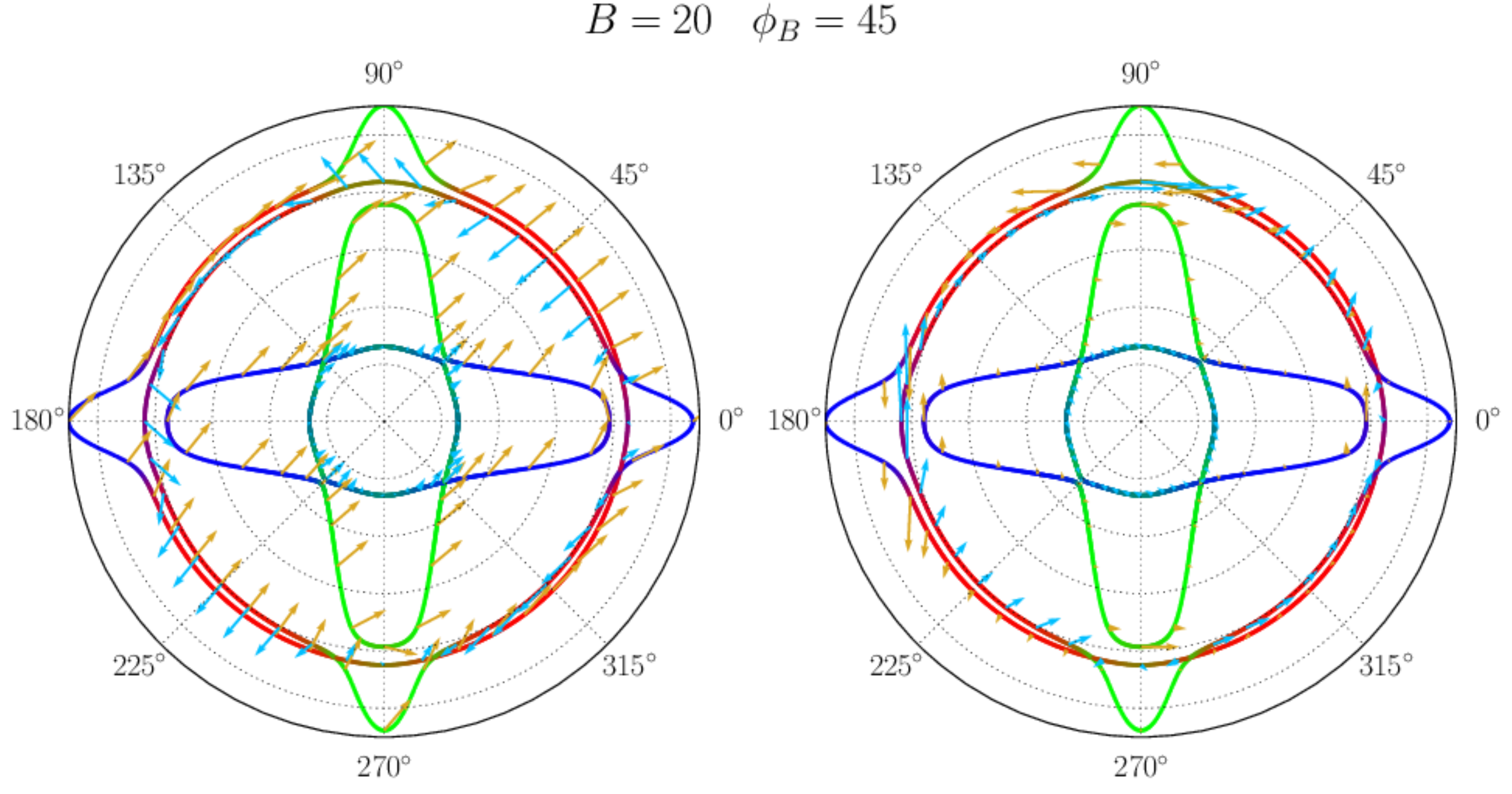}}
\caption{Average spin ({\rm SAM}) and orbital ({\rm OAM}) angular momentum around the Fermi surfaces at $B=0$ (top), $B=20\,{\rm T}$ and $\phi_B=90^{\circ}$ (middle), $B=20\,{\rm T}$ and $\phi_B=45^{\circ}$ (bottom).  $\Delta_{\rm SO}>\Delta_{\rm Z}$ and carrier density $n=2.2\times 10^{-13} {\rm cm}^{-2}$, as the calculations in Sec.~\ref{numerics}.\label{sam_and_oam}}
\end{figure*}
To model the conduction bands at the interface we use a single-electron Hamiltonian\cite{ruhman2014} where the electronic states are derived from the $t_{2g}$ ($d_{xy}$, $d_{xz}$ and $d_{yz}$) orbitals of \ce{Ti}-atoms. Accounting for a total number of six degrees of freedom (three orbitals times two spin components), the translational invariant Hamiltonian in momentum space has a $6\times6$ matrix-representation that is the sum of the four terms in Eq.~\ref{ham}. The kinetic Hamiltonian 
\begin{align}
&  H_{\rm L} =
  \begin{pmatrix}
	  \epsilon_{xy}(k)- \Delta_{\rm E} & 0 & 0 \\
	 0 & \epsilon_{xz}(k)& \delta(k) \\
	 0 & \delta(k) & \epsilon_{yz}(k)
  \end{pmatrix}\otimes \hat{\sigma}_0,
  \label{eq:HL}\\
\epsilon_{xy}(k)&=2t_l(2 -\cos k_x - \cos k_y),\nonumber\\
\epsilon_{xz}(k)&=2t_l(1-\cos k_x) + 2t_h(1-\cos k_y),\\
\epsilon_{yz}(k)&=2t_h(1-\cos k_x) + 2t_l(1-\cos k_y),\nonumber\\
\delta(k)&=2t_d \sin k_x \sin k_y.\nonumber
\end{align}
describes electrons hopping between \ce{Ti}-orbitals on adjacent sites in the interfacial ($xy$) plane. $d_{xy}$ orbitals have all the lobes lying on the $xy$-plane, $x$- and $y$- hopping amplitudes are equivalently described by a single \emph{light} matrix elements $t_l$. Instead $d_{xz}$ and $d_{yz}$ orbitals have both lobes in-plane and in the direction normal to the interface, giving rise to one light and one \emph{heavy} ($t_h<t_l$) matrix element, respectively. $\Delta_{\rm E}$ is the gain in the on-site energy of $d_{xy}$ states confined at the interface compared to the on-site energy of $d_{xz}/d_{yz}$ states. Inter-orbital matrix elements $\propto \sin k_x \sin k_y$ account for $d_{xz}/d_{yz}$ hybridization with a strength $t_d \approx t_h$ (however, this term does not affect at all the results of our calculations).

At the interface the confining electric field along the $z$-direction breaks the inversion-symmetry and activates transitions from $d_{xy}$ orbitals -- which are even under mirror symmetry -- to $d_{xz}$ and $d_{yz}$ orbitals -- odd under mirror symmetry -- on adjacent metal sites. The inversion-breaking term has the form
\begin{align}
  H_{\rm Z} = \Delta_{\rm Z}
  \begin{pmatrix}
	 0 & i\sin k_y & i\sin k_x \\
	 -i\sin k_y & 0 & 0 \\
	 -i\sin k_x & 0 & 0 \\
  \end{pmatrix} \otimes \hat{\sigma}_0.
  \label{eq:HZ}
\end{align}

Atomic spin-orbit coupling is the same as for the bulk \ce{STO} system, that is
\begin{align}
H_{\rm SO} = \frac{\Delta_{\rm SO}}{2} \sum_{i=x,y,z} \sigma_i \otimes L_i =
\frac{\Delta_{\rm SO}}{2}
\begin{pmatrix}
0 & i\hat{\sigma}_x & -i\hat{\sigma}_y \\
-i\hat{\sigma}_x & 0 & i\hat{\sigma}_z \\
i\hat{\sigma}_y & -i\hat{\sigma}_z & 0
\end{pmatrix},
\label{eq:HSO}
\end{align}
with
\begin{align}
 L_x =\hbar
 \begin{psmallmatrix}
	0 & i & 0 \\
	-i & 0 & 0 \\
	0 & 0 & 0 \\
 \end{psmallmatrix},\,
 L_y = \hbar
 \begin{psmallmatrix}
	0 & 0 & -i \\
	0 & 0 & 0 \\
	i & 0 & 0 \\
 \end{psmallmatrix},\,
 L_z = \hbar
 \begin{psmallmatrix}
	0 & 0 & 0 \\
	0 & 0 & i \\
	0 & -i & 0 \\
 \end{psmallmatrix}
  \label{eq:L}
\end{align}
the representations of the components of the orbital angular-momentum. 
Lastly, the Zeeman Hamiltonian $H_{\rm B}=\mu_{\rm B}(\bm L + g\bm S)\cdot\bm B/\hbar$ is
\begin{align}
H_{\rm B} = \mu_{\rm B}
\begin{psmallmatrix}
g({\rm B}_x \hat{\sigma}_x + {\rm B}_y \hat{\sigma}_y)/2 & i{\rm B}_x \hat{\sigma}_0 & -i{\rm B}_y \hat{\sigma}_0 \\
-i{\rm B}_x \hat{\sigma}_0 & g({\rm B}_x \hat{\sigma}_x + {\rm B}_y \hat{\sigma}_y)/2 & 0 \\
i{\rm B}_y \hat{\sigma}_0 & 0 & g({\rm B}_x \hat{\sigma}_x + {\rm B}_y \hat{\sigma}_y)/2
\end{psmallmatrix},
\end{align}
with ${\rm B}_x = |\bm B| \cos \phi_{\rm B},\ {\rm B}_y = |\bm B| \sin \phi_{\rm B}$ and $\bm S = \hbar\hat{\bm\sigma}/2$. 

The expectation-values of the spin and orbital angular momenta are shown (Fig.~\ref{sam_and_oam}) at zero and high  magnetic field for two particular orientations $\phi_B=90^{\circ}$ (largest ${\rm MR}$) and $\phi_B=45^{\circ}$ (approximately largest $\rho_{xy}$ signal) and the set of parameters in the first line of Table~\ref{param_table}.

\section{Dependence of the anisotropy on the parameters of the model}
\label{parameters}

The parameters which define the model of the interface are taken within the ranges that are set by theoretical and experimental results in literature, e.g. first-principles calculations,  ARPES measurements on the surface of \ce{STO} \cite{king2014, santander2011, plumb2014} and more recently available soft-X-ray ARPES on the \ce{LAO}/\ce{STO} interface.\cite{cancellieri2014} Further estimates from transport measurements \cite{benshalom2010, caviglia2010, vanheer2013, fete2014} give more informations at least about the order of magnitude of the energy scales in the system. 
Below we list and discuss the choices of the parameter-values in the paper and show results of additional calculations at different parameters than in the main text, showing a striking stability of the qualitative features of the data in parameter space. (The calculations here are for Gaussian-correlated impurities.) 

\vspace{0.5\baselineskip}
\begin{table}[h]
\centering
\begin{tabular}{|c|c|c|c|c|c|c|c|}
\hline 
& & & & & & &\tabularnewline
{\bf Fig.} &  $t_l$ & $t_h$ & $\Delta_{\rm E}$ & $ \Delta_{\rm SO}$ & $\Delta_{\rm Z}$ & $g$ & $\xi$\tabularnewline
& & & & & & &\tabularnewline
\hline 
& & & & & & &\tabularnewline
{\bf 2 - 3 - 8} & $400$  & $12.5$ & $65$  & $7$ & $2.5$ & $5$ & $5$\tabularnewline
\hline 
& & & & & & &\tabularnewline
{\bf 6} & $400$  & $12.5$ & $65$  & $9$ & $4$ & $-3.4$ & $5$\tabularnewline
\hline 
& & & & & & &\tabularnewline
{\bf 7} & $400$  & $12.5$ & $65$  & $7$ & $2.5$ & $5$ & $4,\,6,\,8$\tabularnewline
\hline 
\end{tabular}
\caption{Parameters used for the calculations in the paper, ordered according to the figures they refer to. }
\label{param_table}
\end{table}

\vspace{0.5\baselineskip}

Hopping elements $t_l,\ t_h$, confinement energy $\Delta_{\rm E}$, atomic-spin orbit strength $ \Delta_{\rm SO}$ and inversion-asymmetry parameter $\Delta_{\rm Z}$ are measured in ${\rm meV}$; the $g$-factor is dimensionless and the disorder correlation-length $\xi$ is measured in units of the lattice constant $a$. The values of the light and heavy mass corresponding to the hopping parameters $t_l$ and $t_h$ are $0.6\,m_e$ and $19\,m_e$ respectively ($m_e$ is the bare electron mass). 

The value of $\Delta_{\rm SO}$ from ab-initio calculations \cite{mattheiss1972} or transport experiments
\cite{benshalom2010, caviglia2010} is estimated in a wide range $10 \div 25 {\rm meV}$. (In a seminal work on Raman scattering for the bulk \ce{STO} system Uwe \emph{et al.} \cite{uwe1985} extracted the value $18\,{\rm meV}$).
Here we consider the values $\Delta_{\rm SO}=7\,{\rm meV}$ and $\Delta_{\rm SO}=9\,{\rm meV}$ (so just below the lower limit of the estimated range) and produce qualitatively similar results for the anisotropy, while at the same time changing also $g$ and $\Delta_{\rm Z}$. In principle one could take larger values of $\Delta_{\rm SO}$ and slightly different hopping elements and still remain in a regime where our results still hold. Moreover, we point out that the strong anisotropy of the spin-orbit field around the Fermi surfaces \cite{king2014} -- with a large enhancement of the effective orbital angular momentum near hybridization gaps -- so far has not been considered in fitting transport measurements, that might return overestimated values of $\Delta_{\rm SO}$.


Outcomes of Boltzmann calculations \cite{diez2015} were found in good agreement with the experimentally measured magnetoresistance at $\phi_B = 90^{\circ}$ (no transverse current, hence $\sigma_{xy}=0$) in the regime of strong inversion-symmetry-breaking $\Delta_{\rm Z}>\Delta_{\rm SO}$. Here we recover a comparable {\rm MR} -- and comparable field-dependence -- in the different (and maybe more realistic) regime $\Delta_{\rm Z}<\Delta_{\rm SO}$. We can understand this similarity observed in two completely different regimes by realizing that the reversal of the spin-ordering on neighbouring Fermi surfaces induced by the magnetic field, which reduces the overall forward scattering and hence lowers the resistance, occurs in both cases regardless of the relative orientation of orbital and spin angular momenta (which is different in the two regimes). 
For $\Delta_{\rm Z}$ we consider values below $5\,{\rm meV}$ which was identified as an upper bound to the real value by Ruhman \emph{et al.}.\cite{ruhman2014}.

\begin{figure}[t]
\begin{center}
\includegraphics[width=0.4\textwidth]{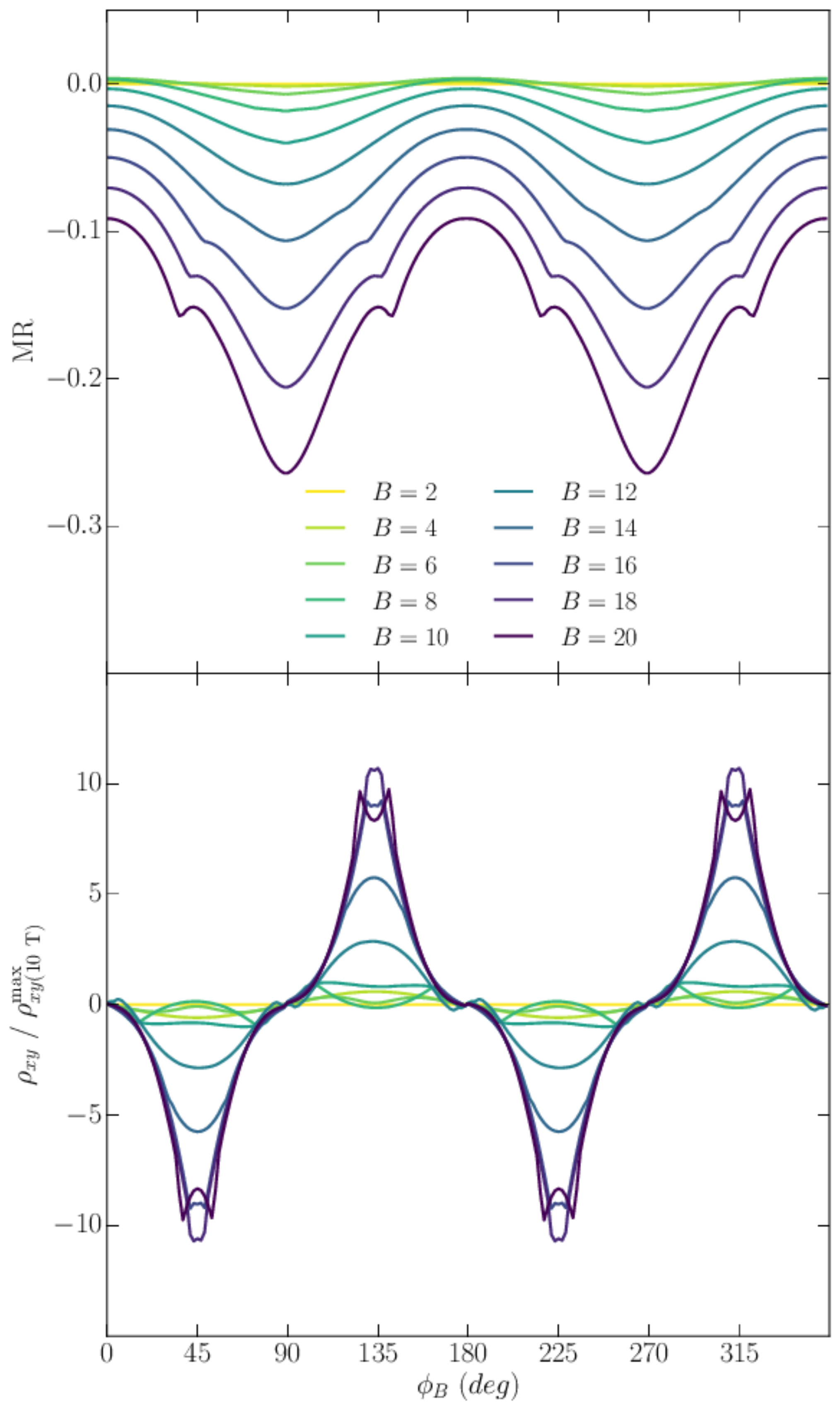}
\caption{${\rm MR}$ and $\rho_{xy} / \rho^{max}_{xy\ (10\,{\rm T})}$ at negative $g$-factor (full set of parameters listed in the second line of Table~\ref{param_table}). \label{g_negative}}
\end{center}
\end{figure}

It is known that the $g$-factor for electrons confined in quantum wells, like \ce{InSb} and \ce{GaAs} \cite{litv2008} can substantially differ from the conventional value $g=2$. In Sec.\ref{numerics} we show results for $g=5$, one of the two possible outcomes (the other one is $g=-3.4$) of a fit to Shubnikov-de Haas oscillations at low temperature\cite{fete2014}. (Note that changing the $g$-factor is not simply equivalent to rescale the magnetic field: magnetic field also couples to the orbital angular-momentum and the relative strength $\langle \bm \mu \cdot \bm B \rangle / \langle \bm L \cdot \bm B \rangle$ is dependent on $g$.) Yet the phenomenology of the anisotropy which we extensively discussed in Sec.~\ref{discussion} is recovered at negative $g=-3.4$ (Fig.~\ref{g_negative}). 

\begin{figure*}[t]
	\begin{center}
		\subfigure[]{\includegraphics[scale=0.28]{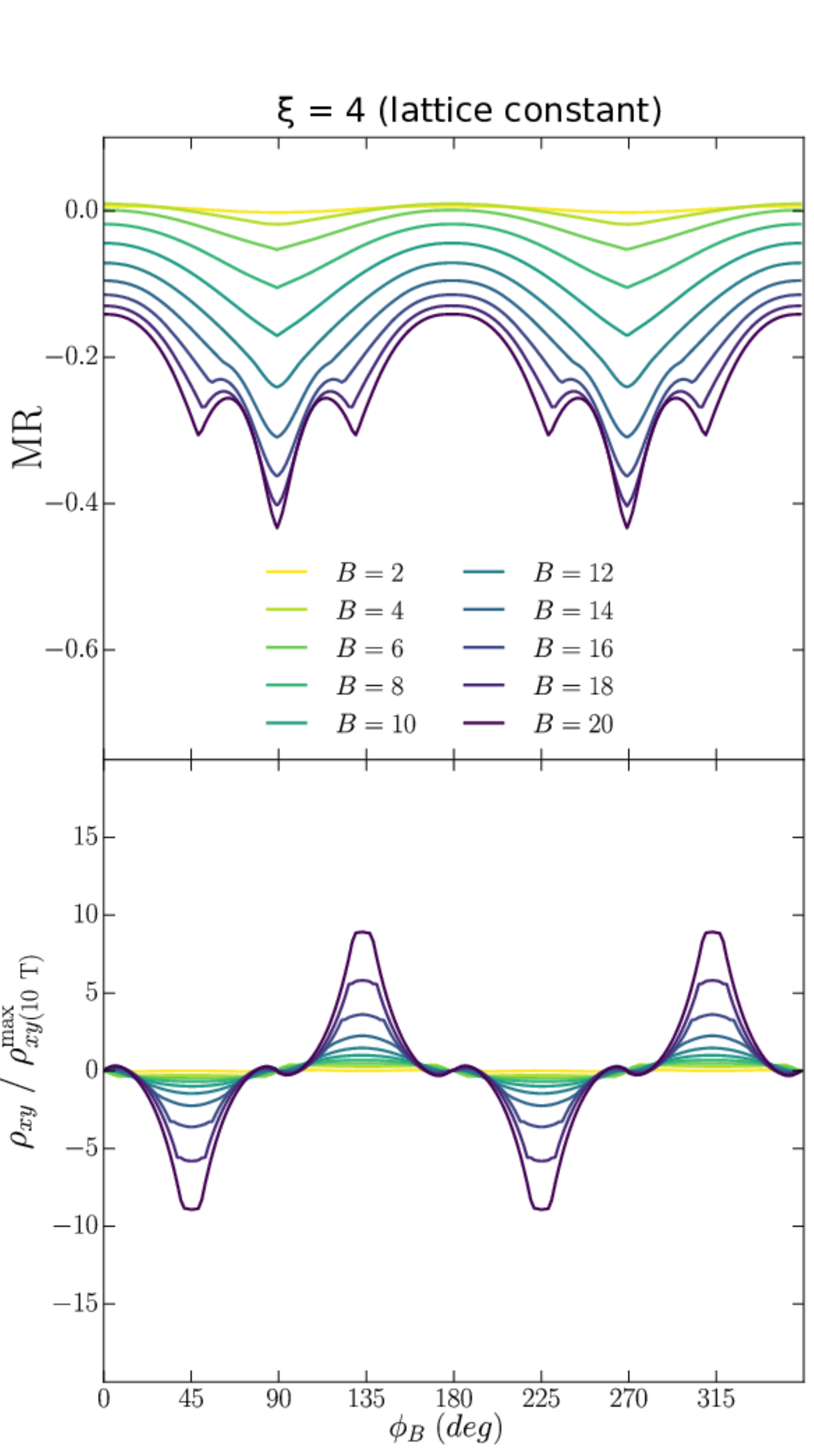}}
		\subfigure[]{\includegraphics[scale=0.28]{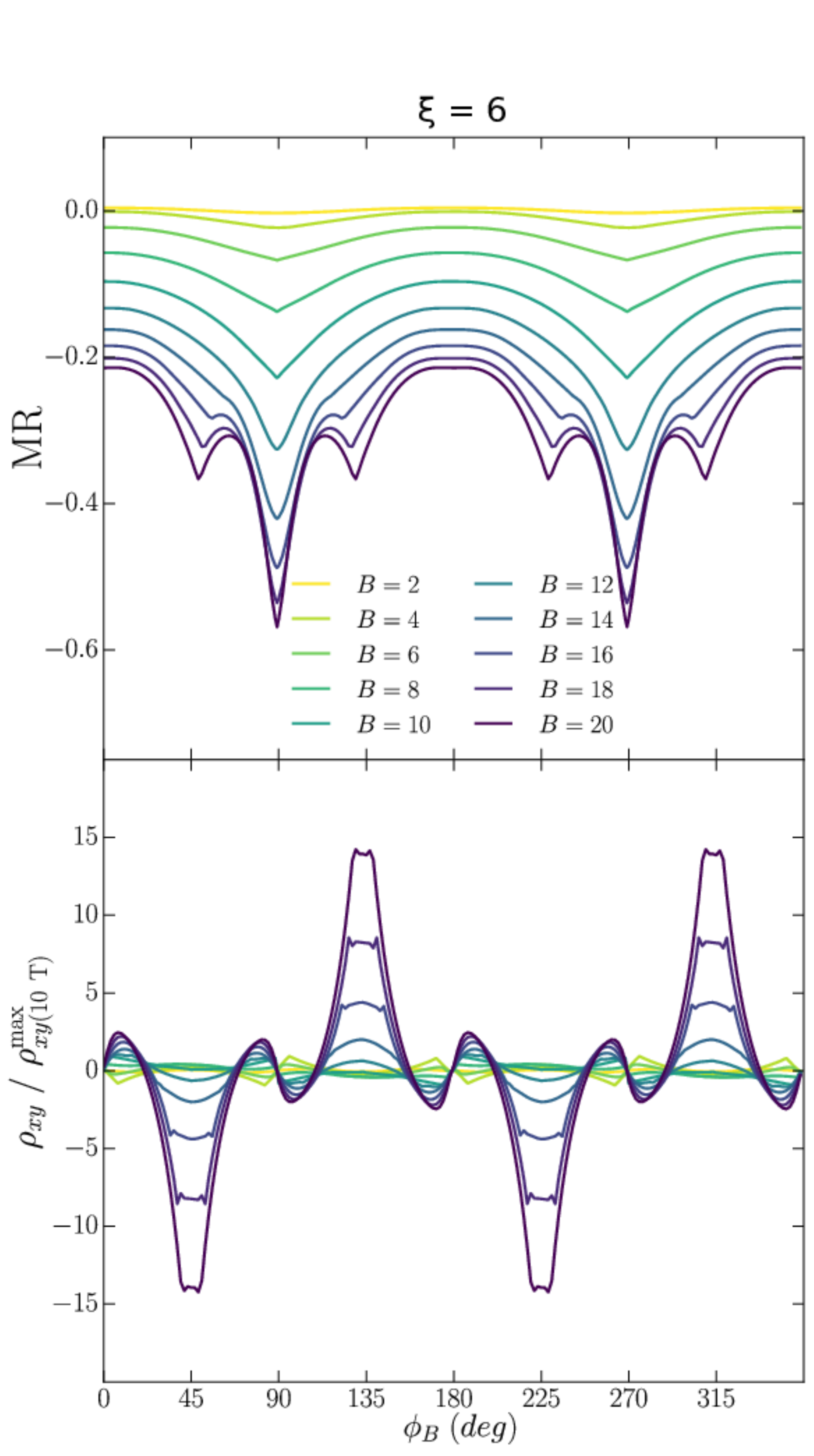}}
		\subfigure[]{\includegraphics[scale=0.28]{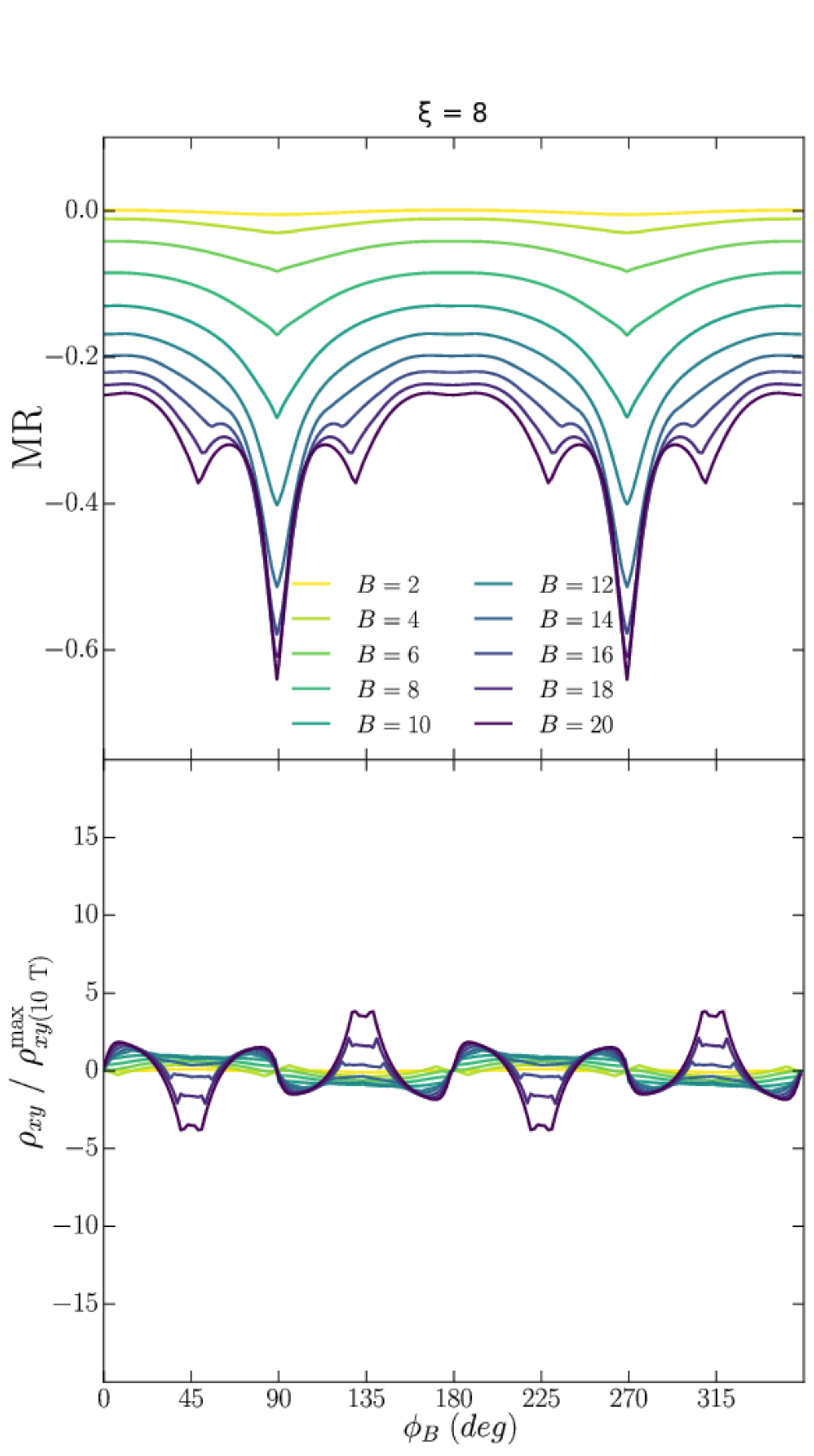}}
		\caption{${\rm MR}$ and $\rho_{xy} / \rho^{max}_{xy\ (10\,{\rm T})}$ at different values of the disorder correlation-length $\xi$ (full set of parameters listed in the third line of Table~\ref{param_table}).\label{more_xi}}
	\end{center}
\end{figure*}

Variations in the correlation-length $\xi$ are also considered. While there is no simple way to extract informations from experiments, it is reasonable to limit $\xi$ within a range of one order of magnitude. Indeed, a too large $\xi$ ($> 10\ a_0$ with $a_0  = 0.4 {\rm nm}$ the lattice constant) would require to treat the impurities as a disordered medium rather than independent scatterers. Calculations in the main text refer to $\xi=5\, a_0$. Below results for $\xi =4,\,6,\,8$ are shown (Fig.~\ref{more_xi}). 
Qualitatively the results are very similar if $\xi|\Delta\k^{bs}|>1$ where $|\Delta\k^{bs}|\sim 2 k_F^{out}$ is the momentum-transfer for backscattering in the large outer band (approximately equal to twice the average Fermi momentum) and at the same time not larger than $10-15\ a_0$ -- with $a_0  = 0.4 {\rm nm}$ the lattice constant -- whereby also the zero-field inter-band scattering is highly reduced. This upper limit is also consistent with the assumption of scattering by individual impurities (rather than by a disordered medium that is a more suitable description for very large $\xi$).
Note that the density of impurities $n_{\rm imp}$ and the amplitude $\delta$ both drop out the expressions for ${\rm MR}$ and $\rho_{xy} / \rho^{max}_{xy\ (10\,{\rm T})}$. 

Finally, let us comment on the density-field dependence of our calculations. A universal scaling of the {\rm MR}-curves as a function of carrier-density, if the magnetic field is rescaled by a density-dependent characteristic value, seems to be a general feature of the experimental data\cite{joshua2013, diez2015}. This is not recovered by the Boltzmann model (even within this different spin-orbit regime) pointing to a physics that might be unrelated to spin-orbit coupling. 

\begin{figure*}
\begin{center}
\subfigure[$\ \ \ \mathsf{n=1.5 \cdot 10^{13} {\rm cm}^{-2}}$]{\includegraphics[width=0.3\textwidth]{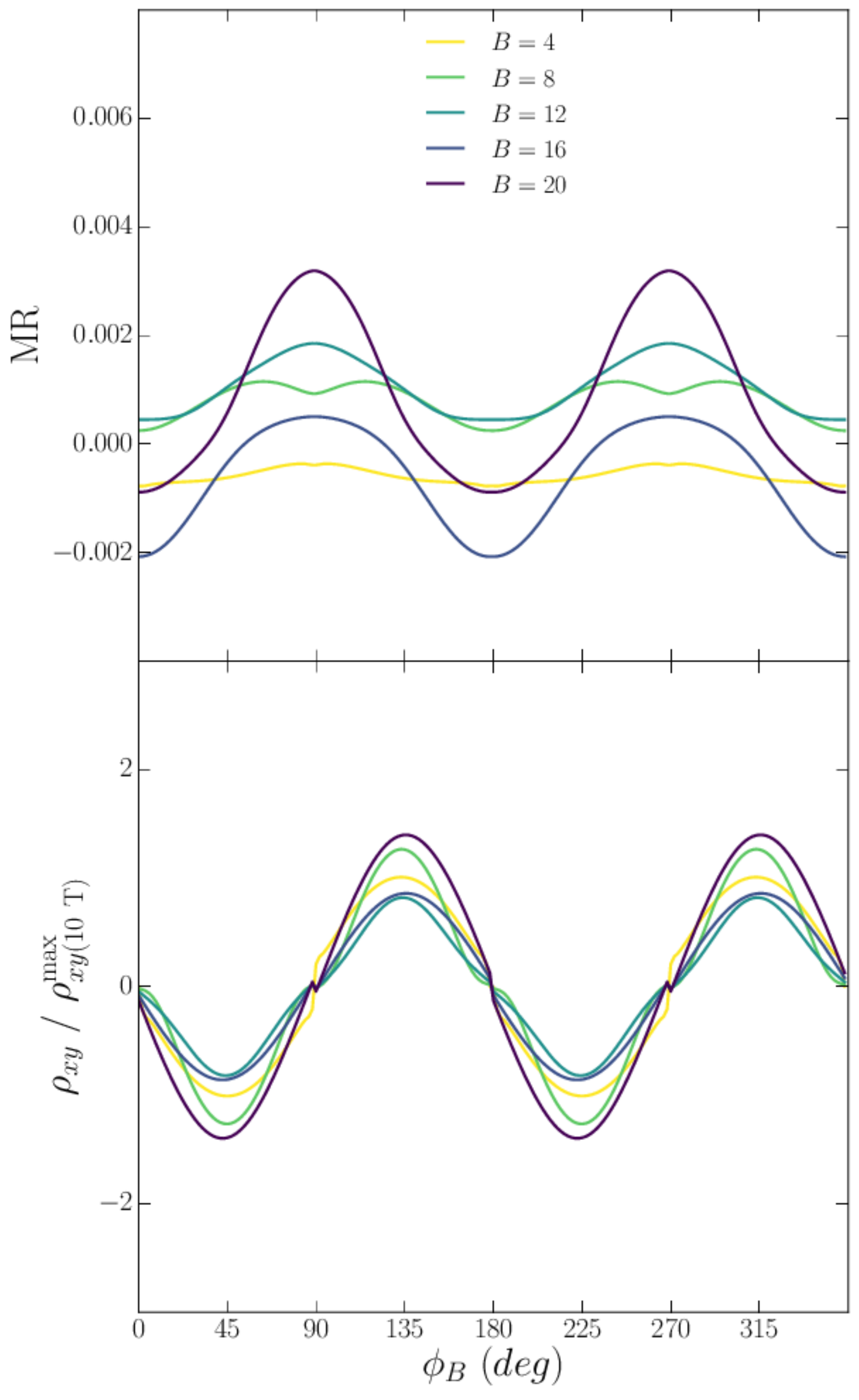}}
\subfigure[$\ \ \ \mathsf{n=2.1 \cdot 10^{13} {\rm cm}^{-2}}$]{\includegraphics[width=0.3\textwidth]{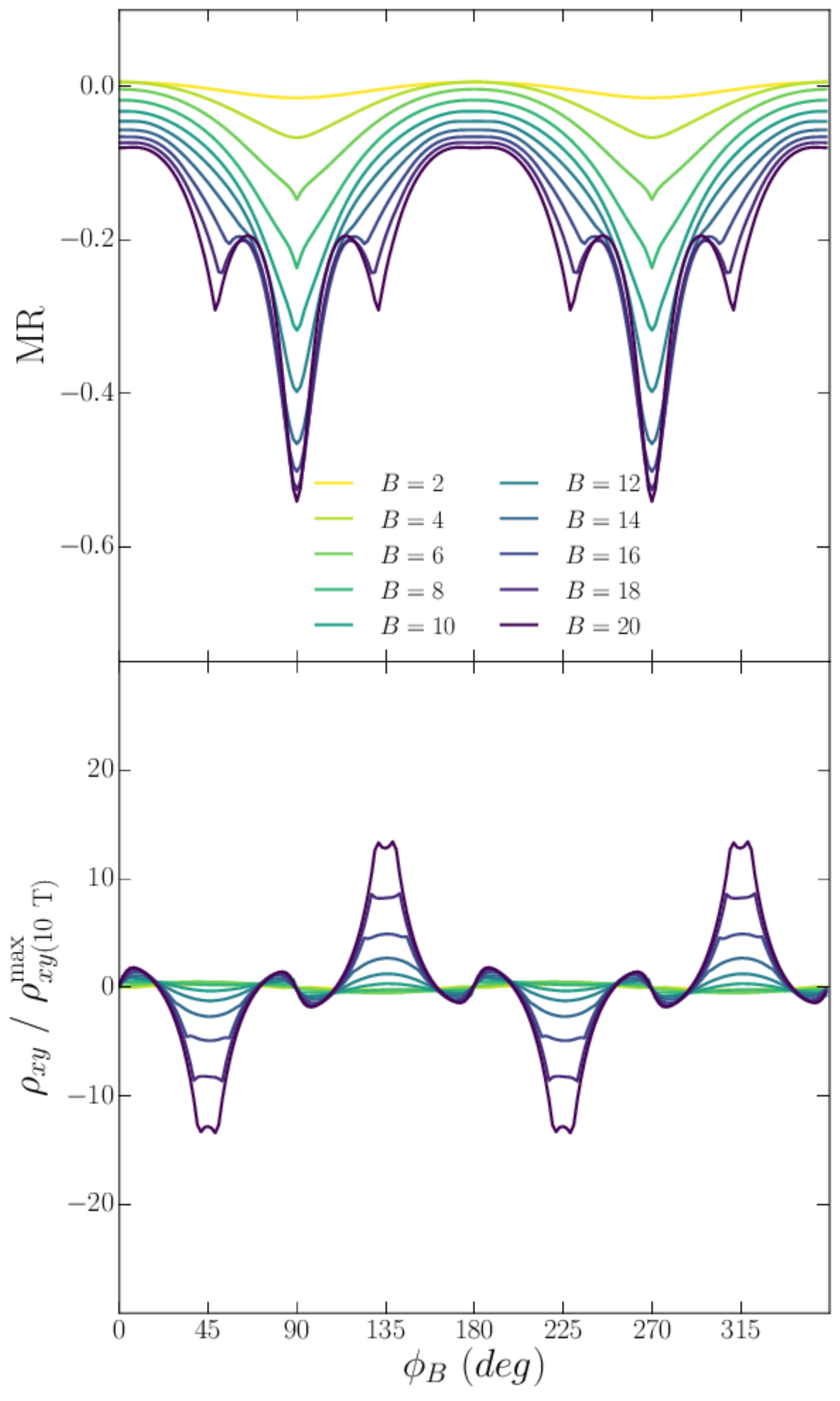}}
\caption{${\rm MR}$ and $\rho_{xy} / \rho^{max}_{xy\ (10\,{\rm T})}$ at lower carrier-densities than the calculations in the main text.\label{lower_densities}}
\end{center}
\end{figure*}

In Fig.~\ref{lower_densities} we show results of calculations at two different densities than the calculations in the main text: $n=1.5 \cdot 10^{13} {\rm cm}^{-2}$ (below the Lifshitz point) and $n=2.1 \cdot 10^{13} {\rm cm}^{-2}$ (above the Lifshitz point). 
The total absence of magnetoresistance at the lowest density (left panel) simply comes from the absence of inter-band scattering (since only the lowest $d_{xy}$ states are filled). 
At higher density (right panel) the high-field {\rm MR} is characterized by multiple maxima and minima as in Fig.~\ref{anisotropy_plot}. It is worth to notice that there is larger discrepancy between ${\rm MR} (\phi_B=0)$ and ${\rm MR} (\phi_B=90)$ than compared to the results at $n=2.2 \cdot 10^{13} {\rm cm}^{-2}$ (Fig.~\ref{anisotropy_plot}). This gap is progressively reduced as the chemical potential is increased up to the middle of the spin-orbit gap at the $\Gamma$-point. 
Densities too close to the Lifshitz point are not considered here. (Remind that the Boltzmann model fails approaching band-edges where $k_F \rightarrow 0$)


\end{document}